\documentclass[conference]{IEEEtran}
\usepackage{url}
\usepackage{amsfonts}
\usepackage{amsthm}
\usepackage{amsmath}
\usepackage{amssymb}
\usepackage{subcaption}
\usepackage{etoolbox}
\usepackage{array}
\usepackage{xstring}
\usepackage{algorithm}
\usepackage{algpseudocode}

\usepackage[inline]{enumitem}
\usepackage{mathtools}
\usepackage{xcolor,soul}
\usepackage{xspace}
\usepackage{cite}

\usepackage{ifthen}

\newboolean{showcomments}
\setboolean{showcomments}{true}
\ifthenelse{\boolean{showcomments}}
{ \newcommand{\mynote}[3]{
    \protect\fbox{\bfseries\sffamily\tiny#1}
    {\small$\blacktriangleright$\textsf{\emph{\color{#3}{#2}}}$\blacktriangleleft$}}}
{ \newcommand{\mynote}[3]{}}

\newcommand\sysname{FITT\xspace}

\newcommand\para[1]{\vspace{0.01in} \noindent \textbf{#1.}}

\def\first{({i})\xspace}
\def\second{({ii})\xspace}
\def\third{({iii})\xspace}

\definecolor{verylightgray}{gray}{0.8}

\usepackage{etoolbox}
\usepackage{xstring}
\DeclareListParser{\doslashlist}{/}
\newcounter{ndnNameComponentCounter}%
\newcommand{\name}[1]{{%
  \setcounter{ndnNameComponentCounter}{0}%
  \renewcommand{\do}[1]{{%
    \ifnumgreater{\value{ndnNameComponentCounter}}{0}{\allowbreak/}{}%
    \ifnumodd{\value{ndnNameComponentCounter}}{}{}%
    \detokenize{##1}}%
    \stepcounter{ndnNameComponentCounter}}%
``{\fontfamily{cmtt}\small\selectfont\IfBeginWith{#1}{/}{/}{}\doslashlist{#1}}''%
}}

\usepackage{etoolbox}
\usepackage{xstring}
\newcommand{\namesm}[1]{{%
  \setcounter{ndnNameComponentCounter}{0}%
  \renewcommand{\do}[1]{{%
    \ifnumgreater{\value{ndnNameComponentCounter}}{0}{\allowbreak/}{}%
    \ifnumodd{\value{ndnNameComponentCounter}}{}{}%
    \detokenize{##1}}%
    \stepcounter{ndnNameComponentCounter}}%
``{\fontfamily{cmtt}\tiny\selectfont\IfBeginWith{#1}{/}{/}{}\doslashlist{#1}}''%
}}

\makeatletter
    \newcommand{\linebreakand}{%
      \end{@IEEEauthorhalign}
      \hfill\mbox{}\par
      \mbox{}\hfill\hspace{-4mm}\begin{@IEEEauthorhalign}
    }
    \makeatother

\begin{document}

\title{Expect More from the Network: \\ DDoS Mitigation by Named Data Networking}
%%\title{Expect More from the Network: \\ DDoS Mitigation by FITT in Named Data Networking}

%% author names and affiliations
%% use a multiple column layout for up to three different
%% affiliations
\author{\IEEEauthorblockN{Zhiyi Zhang}
\IEEEauthorblockA{UCLA\\
zhiyi@cs.ucla.edu}
\and
\IEEEauthorblockN{Vishrant Vasavada}
\IEEEauthorblockA{UCLA\\
vasavada@cs.ucla.edu}
\and
\IEEEauthorblockN{Siva Kesava Reddy Kakarla}
\IEEEauthorblockA{UCLA\\
sivakesava@cs.ucla.edu}
\linebreakand
\IEEEauthorblockN{Angelos Stavrou}
\IEEEauthorblockA{Virginia Tech\\
angelos@vt.edu}
\and
\IEEEauthorblockN{Eric Osterweil}
\IEEEauthorblockA{George Mason University\\
eoster@gmu.edu}
\and
\IEEEauthorblockN{Lixia Zhang}
\IEEEauthorblockA{UCLA\\
lixia@cs.ucla.edu}
}

%\IEEEoverridecommandlockouts
%\makeatletter\def\@IEEEpubidpullup{6.5\baselineskip}\makeatother
%\IEEEpubid{\parbox{\columnwidth}{
%    Network and Distributed Systems Security (NDSS) Symposium 2020\\
%    23-26 February 2020, San Diego, CA, USA\\
%    ISBN 1-891562-61-4\\
%    https://dx.doi.org/10.14722/ndss.2020.23xxx\\
%    www.ndss-symposium.org
%}
%\hspace{\columnsep}\makebox[\columnwidth]{}}

\maketitle

\begin{abstract}

Distributed Denial of Service (DDoS) attacks have plagued the Internet for decades, but the basic defense approaches have not fundamentally changed.
Rather, the size and rate of growth in attacks have actually outpaced carriers' and DDoS mitigation services' growth, calling for new solutions that can be, partially or fully, deployed imminently and exhibit effectiveness.
In this paper, we examine the basic functions in Named Data Networking (NDN), a newly proposed Internet architecture, that can address the principle weaknesses in today's IP networks.
We demonstrate by a new DDoS mitigation solution over NDN, Fine-grained Interest Traffic Throttling \textit{FITT}, that NDN's architectural changes, even when incrementally deployed, can make DDoS attacks fundamentally more difficult to launch and less effective.
FITT leverages the NDN design to enable \emph{the network} to detect DDoS from victim's feedback, throttles DDoS traffic by reverse its exact paths through the network, and enforces control over the misbehaving entities \emph{at their sources}.
%FITT pushes remediation from the service provider that is under attack all the way to the distributed attacking sources without disrupting the traffic to other services.
%%
%% In cases like the Mirai attack, where the cumulative attack traffic from smart IoT devices crippled high-capacity service providers using diverse DDoS Tactics Techniques and Procedures (TTPs), FITT would be able to squelch the attack traffic precisely at its distributed sources, without disrupting other legitimate application traffic running on the same devices.
%% LZ: I commented out Mirai
%%
Our extensive simulation results show that FITT can throttle attack traffic with one-way time delay from the victim to the NDN gateway; 
upon activation, FITT effectively stop attack traffic from impacting benign flows, resulting in over 99\% of packets reaching victims being legitimate ones.
%that over 99\% traffic reaching victims are legitimate packets while 
%benign flows remain unaffected at steady-state. 
We further demonstrate that service providers may implement NDN/FITT on existing CDN nodes as an incrementally deployable solution to effectuate the application-level remediation at the sources, which remains unattainable in today's DDoS mitigation approaches.
\end{abstract}

\section{Introduction}

%\todo{todo}
%\zhiyi{comment sample}
%\lixia{comment sample}
%\sichen{comment sample}
%\angelos{comment sample}
%\eric{comment sample}

% - DDoS is an old problem that has not been solved
Distributed Denial of Service (DDoS) attacks have plagued the Internet for decades and often capitalize on inherent properties of today's TCP/IP networking model~\cite{osterweil202021,ddos-market-size}.
While the Internet's current TCP/IP architecture has achieved unprecedented success, its weakness has also been utilized by attackers to launch DDoS attacks. The ever-increasing size, frequency, and sophistication of DDoS attacks calls for new approaches that can be partially or fully deployed imminently.
We propose that this urgent need may accelerate our consideration of a new Internet architecture, and that evidence exists that there may now be economic \emph{incentives} for large operators to upgrade their existing infrastructures to embrace it.

Indeed, starting with early DDoS examples (e.g., attacks from the Trin00 botnet in 1999~\cite{trin00}) through to recent attacks from the Mirai botnet~\cite{understanding-mirai}, the remediation techniques used in today's Internet suggest that our defensive tactics may not be fundamentally keeping pace with attackers~\cite{osterweil202021}. Rather, with attacking botnet nodes swelling in size to hundreds of thousands, and even millions, attacks have grown large enough that their attack volume rivals provisioned capacity of DDoS mitigation providers.
The Mirai botnet serves as a quintessential example, in that it was used to launch some of the largest DDoS attacks in history, and it did so using compromised devices that primarily included Internet of Things (IoT) devices and household appliances that were both easily discoverable and poorly protected~\cite{understanding-mirai}.
We note that DDoS has evolved to being \emph{more distributed} than ever, and to increasingly using application-level semantics (e.g. reflective amplification attacks using DNS, NTP, memcached, etc.).
On the other hand, service operators, providers, and mitigation services~\cite{akamai-prolexic, neustar-ddos, cloudflare-ddos} have had little recourse but to \emph{centralize defenses} and backhaul
% and disrupt application semantics of malicious traffic\lixia{I dont understand this},
or
% to absorb
black-hole
undisrupted attack traffic (``packet love'') in large DDoS mitigation service networks. These DDoS mitigation approaches haul offending packets deeper into the network and require an ever increasing amount of deep packet inspection and state in terms of flow semantics to filter out attack packets, thus they do not scale well.

Most DDoS attacks that are launched on today's Internet are made possible by utilizing features in the TCP/IP network architecture.
Previous work~\cite{jin2003hop,off-by-default,xiaowei05-dos-limiting,handley04-steps,rossow2014amplification} observed that
DDoS attacks often exploit the following specific properties in IP:
\begin{itemize} [leftmargin=*, itemsep=1pt, parsep=2pt, topsep=4pt]
\item \emph{Push-model Communication}:
Any Internet node can send packets to any other IP address.  This leaves DDoS attack victims with no way to stop the attack traffic.

\item \emph{Destination-based Delivery}:
Packet delivery is solely based on the destination address and there is no source address validation by default. Thus, source IP addresses can easily be misattributed, or spoofed, which is a primary feature used by volumetric reflective amplification DDoS attacks~\cite{osterweil202021}.\footnote{This type of DDoS attack has resulted in the largest attacks seen on the Internet to date.}

\item \emph{Limited Expressiveness in TCP/IP protocol stack}:
IP addresses, even with transport port numbers, cannot expressively describe the semantic characteristics of application-layer traffic,
which makes it difficult for DDoS defense mechanisms to inspect traffic to identify attack packets.
%Also, congestion control in TCP/IP architecture is end-to-end and compliance with such controls is dependent on the end host.
\end{itemize}
Moreover,~\cite{handley04-steps} proposed that a DDoS resilience architectural would need:
\begin{enumerate*} [label=(\roman*)]
	\item limiting the access to a server based on the server's capabilities,
	\item source address authentication to prevent source address spoofing,
	\item separating client and server address space to prevent unwanted traffic from client to client and server to server, and
	\item building symmetric traffic flows to prevent reflection attacks at the network layer.
\end{enumerate*}
% Prior work~\cite{handley04-steps}  proposed modifications to the existing Internet architecture to build greater DDoS resilience at an architectural level.
% They identified the following desired features of a DDoS-resistant Internet architecture:
% \begin{enumerate*} [label=(\roman*)]
% 	\item limiting the access to a server based on the server's capabilities,
% 	\item source address authentication to prevent source address spoofing,
% 	\item separating client and server address space to prevent unwanted traffic from client to client and server to server, and
% 	\item building symmetric traffic flows to prevent reflection attacks at the network layer.
% \end{enumerate*}
% % It is unfortunate that realizing these identified features is difficult, if not impossible, in deployed TCP/IP networks.
% % The current state of DDoS mitigation suggests that there is a fundamental impedence to realizing these features in the TCP/IP architecture.
% % In this paper, our examination of NDN in Section~\ref{sec:ndn-ddos} shows that NDN's architecture natively embodies these notions.

In addition to these observations from the literature,
% , two of the biggest challenges in DDoS mitigation challenges in DDoS mitigation
two other challenges facing the DDoS mitigation industry today
are the need for distributed enforcement closer to the adversary, and the ability to utilize existing core network infrastructure by providing an incremental deployment.
Thus, in efforts to meet the distributed DDoS threat with distributed remediation in incrementally deployable ways, approaches like BGP's FlowSpec~\cite{RFC5575}, Remote Triggered Back-Holing (RTBH)~\cite{RFC5635}, among others, have attempted to coordinate defenses by setting up network-level traffic filtering state.
Unfortunately, DDoS Tactics, Techniques, and Procedures (TTPs) are sufficiently nuanced as to need Deep Packet Inspection (DPI), and network-level remediation lacks the necessary expressiveness and requires maintaining distributed state to encode the TTPs.  This has, therefore, led to collateral damage or a lack of adoption of these protocols and techniques.
%%%%%
We argue that the stagnant progress in DDoS mitigation deployment suggests that there may be a fundamental impedance to realizing these features in the TCP/IP architecture.

%\begin{figure}[t]
%	\centering
%	\includegraphics[width=0.45\textwidth]{figures/ndn-ip}
%	\caption{Architectural comparison of TCP/IP vs NDN}
%	\label{fig:ndn-arch}
%	\vspace{-5mm}
%\end{figure}

The insights presented in this paper not only align well with the above literature's prescriptions,
but go farther by offering an architecture, a candidate DDoS mitigation design, \emph{and} an incremental deployment model that aligns deployment costs with incentives and is performance effective.
To that end, in Section~\ref{sec:ndn-ddos} we describe how the Named Data Networking (NDN)~\cite{zhang2014named} architecture (while beneficial in many ways) is principally well suited to addressing the fundamental vulnerabilities that allow today's DDoS attacks.
%we extend Named Data Networking (NDN)~\cite{zhang2014named} (Figure~\ref{fig:ndn-arch}) to investigate whether DDoS defense can benefit from \emph{its} architectural changes.  In Section~\ref{sec:ndn-ddos}, we describe how NDN's architecture (while beneficial in many ways) is principally well suited to addressing the fundamental vulnerabilities that allow today's DDoS attacks.
Indeed, NDN changes the basic network communication model and directly brings application-layer data names to the network layer: instead of pushing packets to IP addresses, an NDN network lets users request named \emph{Data packets} by sending \emph{Interest packets} that carry the desired \emph{data name}.
The network forwarders record the state of Interest packets, making breadcrumb traces for the returning data packets.
%%This stateful forwarding provides abundant traffic insights.

%\begin{figure}[h]
%	\centering
%	\includegraphics[width=0.35\textwidth]{figures/packet}
%	\caption{NDN Interest packet and Data packet}
%	\label{fig:ndn-packet}
%	\vspace{-5mm}
%\end{figure}

We show in Section~\ref{sec:design} that these two key design features provide a solid foundation for an effective DDoS mitigation design: although NDN's receiver-driven model eliminates attacks by data packets, attackers may attack a target in an NDN network by flooding Interest packets.
Leveraging NDN's semantic names at the network layer and stateful forwarding, we develop a novel DDoS mitigation solution -- Fine-grained Interest Traffic Throttling (\textit{\sysname}) to combat Interest DDoS attacks.
\textit{\sysname} enables victims to selectively push back incoming Interest packets towards end-points that are part of an NDN enclave, blocking DDoS attack traffic close to the source.

In Section~\ref{sec:cdn}, we detail observations that large Internet infrastructures, Content Distribution Networks (CDNs),
% already exists to support NDN's rollout. Further, we propose that CDN owners are in a position that is incentivized to adapt the deployments of NDN to benefit their existing business interests.
are already aligned to gain economic benefits from rolling NDN out on their existing distributed infrastructures (synergizing CDN and DDoS enhancements in the network infrastructure).
We show that our approach can be effective with incremental rollout, leveraging the presence of CDNs acting as an overlay connecting up NDN enclaves at network edges. We argue that the architectural changes of the network, even with incremental deployment at the edges, where DDoS originates from, can make DDoS attacks fundamentally more difficult to launch and less effective.

Together with the benefits, an incremental deployment of a new architecture will necessarily face growing-pains.
Among the considerations will be that clients and services will need to behave as NDN edges.
%% operate in a mode that enables both TCP/IP and NDN connectivity.
For edge networks that embrace the architectural benefits of NDN with \sysname while running TCP/IP-based applications, Application-Level Gateways (ALGs) will be needed to provide interoperability mapping and encapsulating of TCP/IP over NDN transport.  While prior work has explored challenges that exist in such approaches\cite{nCDN}, our analyses illustrate that there are fundamental advantages (performance and economic) to be gained by deploying NDN (even as ALGs to bridge with TCP/IP) on both the client and service sides.
Moreover, previous work also raised concerns that NDN's use of semantically meaningful names may lead to privacy leaks. We agree these are relevant trade-offs to be discussed and the issues mitigated, but focus this
work on the potential benefits NDN and \sysname can offer to the DDoS war.

\sysname achieves the following design goals and novel contributions:
\begin{itemize} [leftmargin=*, itemsep=1pt, parsep=2pt, topsep=4pt]
\item To the best of our knowledge, this paper is the first comprehensive description of how NDN's architectural design decisions lead to an inherently resilient foundation for DDoS defense.

\item Building upon the design advantages of NDN our approach, \sysname, offers target-controlled DDoS mitigation using feedback from \emph{the victim}.

\item \sysname responds automatically and swiftly to attacks: it takes a mere half round-trip time between the victim and client-side edge NDN routers for \sysname to detect and start throttling the offending flows mitigating attacks without depending on any human intervention.

\item \sysname supports fine-grained traffic throttling of specific attacking traffic flows close to the attack origin without affecting benign flows when \sysname reaches steady-state.

\item At the edge routers, \sysname introduces reinforcement control to distinguish attack traffic achieving $>$99\% throughput for legitimate traffic.

%% \item We show that \sysname can be deployed at the edges cost effectively using existing CDN infrastructure as an overlay to connect NDN enclaves requiring approximately 20\% of CPU and memory compared to vanilla CDN caching.
\item We show that
% \sysname can be deployed at the edges cost effectively using
existing CDN infrastructure providers may now have an immediate economic incentive to adjust their networks to deploy \sysname.
% , and thereby act as overlays that can connect NDN enclaves.
We show evidence that (in addition to the fundamental DDoS advantage) such an adjustment would also reduce CPU and memory to approximately ${1}/{5}$
% approximately 20\% of CPU and memory
compared to vanilla CDN caching.

%\lixia{would be better to show some numbers}

%\item Our evaluation shows that \sysname is more accurate than existing solutions for DDoS mitigations on TCP/IP networks and other proposed countermeasures for NDN architectures.
\end{itemize}

Our analysis, prototype implementation, and evaluation illustrates that the network can do more to protect application services than we currently expect. By reconsidering the architectural design, we can augment network security in a fundamental way: \textit{let applications instruct the network to squelch DDoS at its sources}. Through our evaluation, we show the NDN/\sysname system only requires an estimate of 20\% for both memory and computational resources compared to traditional CDN caching. When it comes to network bandwidth, NDN incurs a mere 36\% bandwidth (almost one packet for every three required by TCP/IP) than current popular CDN deployment to serve the same number of requests. Moreover, multiple simulations of our NDN/FITT prototype when deployed over CDN shows that FITT reacts effectively to throttle attack traffic requiring half round-trip time between the victim and the client-side NDN gateway to respond to attacks which is usually measured in milliseconds. %Upon activation, \sysname guarantees that over 99\% traffic reaching victims are legitimate packets while benign flows remain unaffected when the system reaches steady-state. \angelos{Please check this statement for validity.}

\section{NDN's Properties for DDoS Mitigation: A Comprehensive Analysis}
\label{sec:ndn-ddos}

\subsection{Named Data Networking}
\label{sec:ndn-basic}

\begin{figure}[t]
	\centering
	\includegraphics[width=0.43\textwidth]{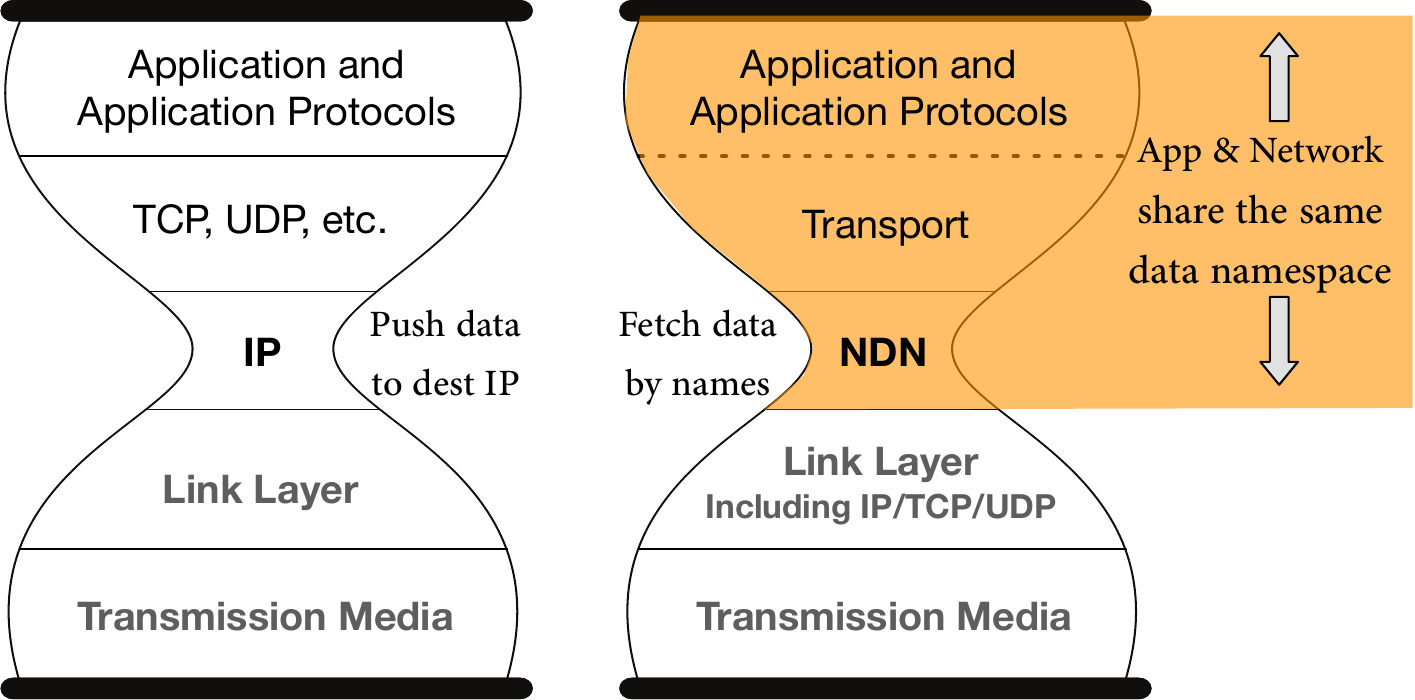}
	\caption{Architectural comparison of TCP/IP vs NDN}
	\label{fig:ndn-arch}
	\vspace{-2mm}
\end{figure}

\begin{figure}[t]
	\centering
	\includegraphics[width=0.43\textwidth]{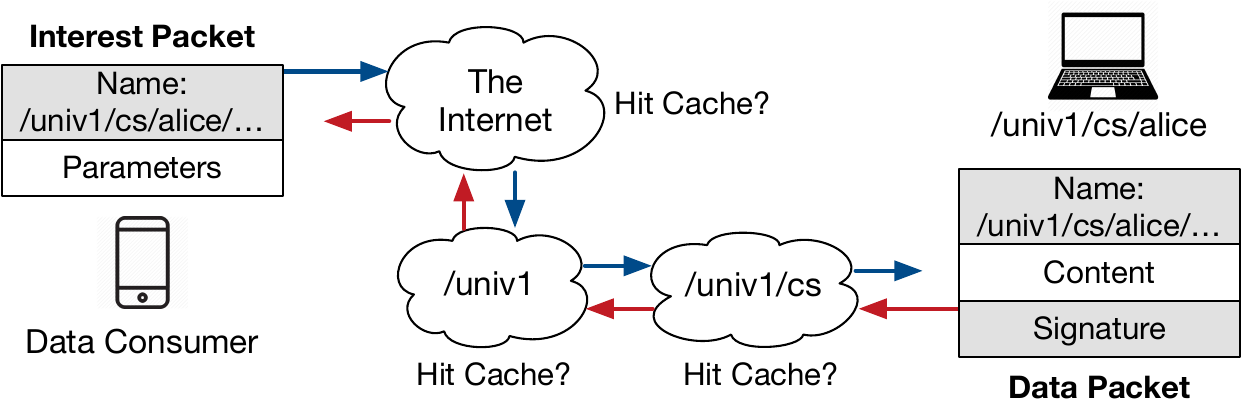}
	\vspace{4pt}
	{\footnotesize \\
		The Interest has a name ``/univ1/cs/alice/video/demo.mp4".
		Each forwarder along the path forwards the Interest packet based on the forwarding information table (FIB) using the longest prefix match. \par}
	\caption{Data Fetching with an Interest}
	\label{fig:prefix-forwarding}
	\vspace{-5mm}
\end{figure}

\para{Named Data Centric}
Named Data Networking (NDN) makes named data the thin waist of the network architecture (Figure~\ref{fig:ndn-arch}).
More specifically, applications name their data at the application layer and NDN directly uses the namespace of applications for network layer data delivery.
These data names are semantically meaningful and structured, e.g., a video produced by Alice's device may have the name \name{/univ1/cs/alice/video.mp4}.
In NDN, routing and forwarding the packets are based on \textit{name prefixes}.
Figure~\ref{fig:prefix-forwarding} shows a simple illustration of NDN Interest-Data exchange.

\para{Build-in Security Building Blocks}
NDN builds communication security~\cite{securityTR} into the architecture by requiring data producers to cryptographically sign all data packets at the time of production and, if needed for content confidentiality, encrypt them as well. Securing data packets directly enables routers to cache them as they pass along, and enables consumers to validate Data packets regardless of where and how they are fetched. Moreover, NDN's routing system~\cite{hoque2013nlsr}, which is based on NDN's Interest-Data exchange, is also secured such that only an authorized user can register its prefix to a forwarder and the communication between routers is also protected.

\para{Stateful Forwarding}
NDN utilizes a stateful forwarding plane: forwarders will record each Interest packet toward data producer, and the fetched Data packet will strictly follow, in reverse, the path taken by the corresponding Interest to get back to the requesting entity. Since an NDN network concerns about data instead of locations, multiple Interest packets requesting the same Data packet are merged in the network (called Interest Aggregation). NDN's forwarding module realizes stateful forwarding by introducing a Pending Interest Table (PIT) into each router. The PIT stores currently unsatisfied (pending) Interests together with their incoming/outgoing interfaces. When a Data packet arrives, the router sends the Data packet to all incoming interfaces recorded in the corresponding PIT entry and removes this PIT entry; the replied Data can be cached in the router's Content Store (CS) to satisfy future Interests requesting the same piece of data. In addition to Interest and Data packets, either NDN routers or data producers may generate NACK packets, which serve as a hop-by-hop feedback mechanism to report a problem in further forwarding of Interests. When a router receives such a NACK packet, it takes appropriate action(s) based on reason code carried in the NACK packet.

\begin{figure} [t]
	\centering
	\includegraphics[width=1\linewidth]{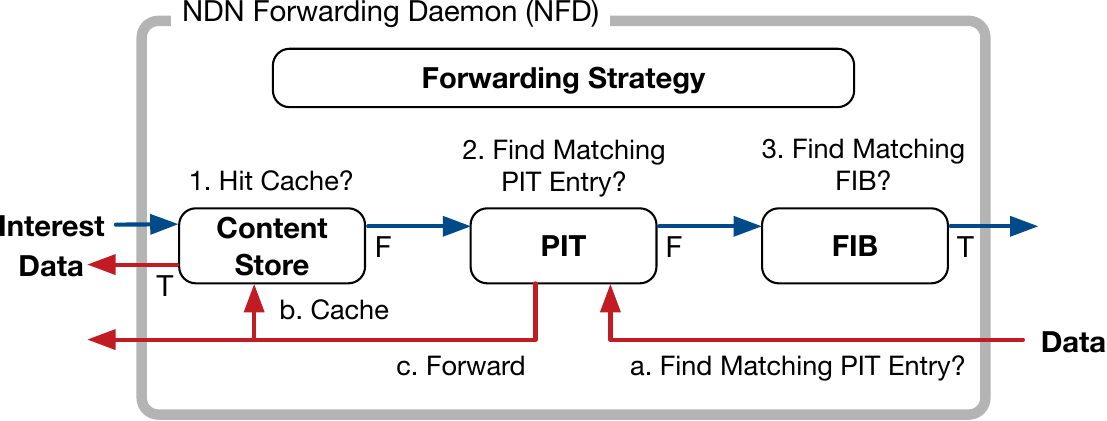}
	\caption{NDN Forwarding}
	\label{fig:ndn-forwarding}
	\vspace{-4mm}
\end{figure}

\subsection{NDN's DDoS Mitigation Properties}

In this section, we examine NDN's architectural advantages over DDoS defenses in the TCP/IP architecture in terms of DDoS resiliency.
%Through the examination, we show NDN's architecture, by its design, leads to DDoS resilience as follows:
We identify the following advantages when NDN is deployed:
\begin{enumerate*} [label=(\roman*)]
	\item NDN's Interest-Data packet exchange eliminates network-layer reflection attacks and DDoS attacks by flooding Data packets.
	\item The data pull model and securing data directly make it more difficult for attackers to recruit ``zombie armies'' because there are no open interfaces or network ports to "scan".
%	\item NDN can mitigate DDoS by route hijacking and cache poisoning by its secured routing system.
	\item By design, NDN caches content automatically using Interest aggregation and thus pushed the overload caused by DDoS away from the victim for static and existent Data.
	\item NDN's stateful forwarding offers rich insights into ongoing traffic for DDoS defense mechanisms.
\end{enumerate*}

\para{Traffic is Off By Design}
\label{sec:ndn:pull}
For devices that serve content in TCP/IP servers, DDoS threats begin as soon as a service goes online and becomes instantly reachable globally. By contrast, the communication in NDN follows a pull model and an application or a node is considered as ``off by design"~\cite{off-by-default} for the following reasons:
\begin{enumerate*}
\item One cannot send an Interest to a consumer application, or a producer application whose name is unreachable from the sender.
By simply not announcing its prefix to solicit Interest packets, an application can still pull data from others, but can \emph{not} be reached by an Interest packet, thus reducing the attack surface for malcode infections and DDoS attackers.

\item A Data packet cannot go anywhere if it is not requested, because there is no corresponding Interest path.

\item Flow-parity: one Interest can at most bring one Data packet back.
\end{enumerate*}
With the pull model, an attacker cannot launch DDoS by flooding Data packets, thus network layer DDoS attacks can only be carried out by Interest flooding.
However, unlike TCP/IP,
as Data Packets follow the reverse path of their corresponding Interests, an attacker cannot redirect them to another consumer.
By design NDN \emph{fundamentally} eliminates reflection DDoS attacks at the network layer.
%% EO: Please don't use this acronym.
%% and distributed reflection DoS (DrDoS) attack.

\para{Data Name Instead of IP Address}
In TCP/IP, even clients who do not run services can be attacked, compromised, and enslaved.
One very common intrusion TTP of attackers is to scan the IP address
space in order to discover devices, and then compromise them.
In NDN, however, if an end device does not \emph{serve} data, it does not even need a name and can eliminate the attack surface of being exposed at all.
More so, routers in NDN forward an Interest by its name.
%% Unlike IP address which is numerical and of a fixed format, an
NDN names are defined by the application semantics in an arbitrary format that are not enumerable.
%When a producer's name prefix is not widely known (e.g., only known to local network), it is difficult for an outside attacker to ``guess'' the exact prefix owned by the target host.
For example, a smart home device with an application-defined name \name{/my/name/home/refrigerator-02} is less exposed compared with an IP address with a default port number, like Mirai
exploits~\cite{understanding-mirai}, because in NDN the exposed network prefix may only be \name{/my/name} and ``guessing'' the exact name requires reconnaissance.
If an Interest name does not match a specific prefix in the forwarding table, the packet will get dropped by the router.
Using application-defined names fundamentally makes a source more difficult to be found and then compromised.

Another big benefit of using a name is to allow the network to inspect traffic at a much finer granularity.
For example, a compromised smart home refrigerator, in a Mirai botnet, may be carrying out network transactions with its device-manufacturer while \emph{also} being forced to participate in a DDoS attack.
There, the legitimate Interest traffic might have prefix \name{/iot-provider/service} and attacking traffic might be towards prefix \name{/com/target/service1}.
Since names of data are directly exposed to the network, the infrastructure is able to identify specific application-level traffic flows and squelch just them
(and not the legitimate traffic).

\para{In-network Cache and Interest Aggregation}
\label{sec:cache}
\label{sec:interest-aggregation}
NDN's content-centric communication model provides enhanced data availability by enabling caching inside the network (i.e. the CS in NDN Forwarders).
Because of the in-network caching, Data packets carrying static content (e.g. HTML files, CSS files, images) can be cached by routers to satisfy future Interest packets, thus reducing the number of Interests reaching the producer (victim). In addition, Interests targeting the same piece of the named data will be aggregated by the router and later Interest packets will not be sent out.
This feature makes it harder for DDoS attackers to flood the same Interest packet or a small set of Interest packets towards the producer in a short time.

As shown in previous work~\cite{psaras2012probabilistic} and our simulation results in Section~\ref{sec:evaluation:agg-cache}, in-network caching and Interest aggregation help to mitigate the Interest flooding where attackers send Interests for static or existing Data packets.
However, if attackers flood a target prefix with a large set of Interest packets or even fake Interests with randomly generated components, the benefits will diminish.  This is because churn and evictions in the cache will lead to diminishing cache hit-rates and the chance of two attacking Interests sharing the same name drops.
%However, since caches have limited capacity, when attackers are able to send Interests for a large number of Data packets, the benefits of caching is reduced because the chance for an Interest to hit a cached Data is lower.
However, it is also noteworthy that NDN is incrementally deployable and does not immediately require rich deployment and caching in the routed core of networks in order to function properly.

%\subsubsection{Reduce Attack Traffic by Interest Aggregation}
%In NDN, Interests targeting the same piece of the named data will be aggregated by the router and later Interest packets will not be sent out as discussed in section \ref{sec:ndn-basic}.
%When the corresponding Data is fetched, the router will send a copy to each incoming interface as recorded in the PIT entry.
%This feature makes it harder for DDoS attackers to flood the same Interest packet or a small set of Interest packets towards the producer.
%However, if attackers flood a target prefix with a large set of Interest packets or even fake Interests with randomly generated components, the benefits of Interest aggregation will diminish because churn and evictions in the cache will lend to diminishing cache hit-rates.

\para{Rich Traffic Insight by Stateful Forwarding}
As NDN's deployment pervades more of the routing infrastructure, its
% NDN's
stateful forwarding~\cite{yi2013case} provides rich insight into ongoing traffic.
Different from a router in TCP/IP, which has little knowledge about which downstream interface attackers are behind, stateful forwarding in NDN helps forwarders to know exactly which interface the traffic is coming in from, by design.  This helps NDN traceback to misbehaving clients and reinforces mitigation.
By observing each Data packet and its corresponding pending Interest entry in the PIT, an NDN forwarder is able to measure the round-trip time, throughput, and name reachability of each outgoing interface.
Previous work\cite{afanasyev2013interest, compagno2013poseidon} shows that a forwarder can also learn the Interest satisfaction ratio, namely the proportion of Interests that successfully fetched a Data packet, and thus detect possible fake Interest DDoS attacks.
Moreover, PIT entry timeouts also offer relatively cheap DDoS attack detection~\cite{zhang2014named}.

\section{The Design and Characteristics of \sysname}
\label{sec:design}

%% <EO>  I moved this from the NDN section (for now)
\subsection{Protecting NDN against remaining DDoS attacks}
\label{sec:threat}

In NDN, attackers can only attempt to DDoS a target by flooding it with Interests.
Specifically, inspired by the work~\cite{Gasti2013DoSD}, we categorize Interest packets used in DDoS attacks into three types according to \first whether the Interest is valid, and if valid, \second whether or not the data requested by the Interest is dynamic (i.e., generated upon the Interest).

\para{Valid Interest for static data (Valid-S)}
Valid-S Interests fetches Data packets that can be cached, e.g., Data packets for a CSS file or a video chunk.
In NDN, Valid-S attacks can be mitigated because multiple Interests asking for the same data can be aggregated and satisfied by the in-network cache.
NDN's intrinsic mitigation is sufficient unless attackers flood a huge amount of Interests traffic from a large spectrum of names.

\para{Valid Interests for dynamic data (Valid-D)}
Valid-D Interests request data that is dynamically generated by producers upon the arrival of the Interest packets, for example, Interest packets used in a remote procedure call.
This is usually reflected by a dynamic and unique data name carried by Interest packets.
Since the Interest's name is customized and the Data is generated in real time, hardly any Interests arriving at a forwarder would hit cache or an existing Interest with the same name.

\para{Invalid Interests (Invalid)}
An invalid Interest packet will not fetch Data packet back because of its unrecognized format, unverifiable signature, incorrect application-layer content, etc.
This type of Interest is mostly useful to malicious adversaries because legitimate applications can generate correct Interests following certain naming conventions.
As discussed in \cite{afanasyev2013interest, compagno2013poseidon}, a possible way to generate such Interests is to append non-existent name components (e.g., randomly-generated garbled bytes) to valid server prefixes.

When Valid-D and Invalid Interests are used in a DDoS attack, since their names are arbitrary and can hardly be satisfied by cache, additional DDoS mitigation services are needed over NDN.

\subsection{\sysname System Model}\label{sec:subdesign}

\sysname is presented to address the attack surface of Interest flooding by enabling data producers to identify and push back fine-grained remediation to attack sources.

%\begin{figure}[t]
%	\centering
%	\includegraphics[width=0.35\textwidth]{figures/toy}
%	\caption{An example topology. Each link can be an overlaid NDN tunnel over TCP/IP network.}
%	\label{fig:toy}
%	\vspace{-3mm}
%\end{figure}

\begin{figure}[t]
	\centering
	\includegraphics[width=0.35\textwidth]{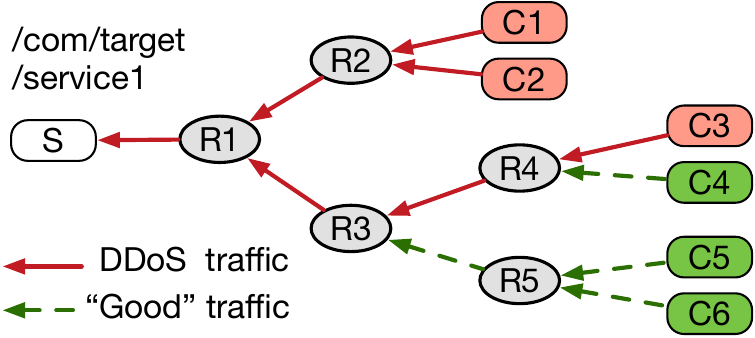}
	\caption{An example topology. Each link can be an overlaid NDN tunnel over TCP/IP network.}
	\label{fig:toy}
	\vspace{-3mm}
\end{figure}

We start by introducing an example topology shown in Figure~\ref{fig:toy}:
The \emph{server $S$} runs an NDN service, producing data under the prefix $P$ \name{/com/target/service1}.
\emph{$C1$-$C6$} are NDN clients requesting data from $S$, in which $C1, C2, C3$ represent compromised devices from a botnet and flood Interest packets.
The	\emph{routers $R1$ to $R5$} are forwarders that support NDN's network stack, e.g., NDN overlay deployed on CDN point of presences.
For the sake of explanation, hereafter, we call the routers towards the server ``upstream routers'' and routers towards the clients ``downstream routers'', e.g., $R4$ and $R5$ are downstream routers of $R3$.

\para{Goals}
\sysname is designed to achieve the following goals.
\begin{enumerate} [leftmargin=*, itemsep=1pt, parsep=2pt, topsep=4pt]
	\item Generality. Our proposed system aims to defend not only from Invalid Interest DDoS attacks, but also from Interest flooding with Valid-D and Valid-S Interest packets, and mixed attack with all three types of Interests.

	\item Fine granularity. \sysname should mitigate Interest flooding traffic precisely.
	To be more specific, traffic from legitimate clients and the traffic towards non-victim services should not be affected.
	For example, when $S$'s service \name{/com/target/service1} is under attack, our proposed approach should throttle away the DDoS traffic towards this service only.
	At the same time, legitimate traffic towards $S$'s other services (e.g., \name{/com/target/service2}) and other servers should not be affected even if the traffic is from attacking clients. In this way, collateral damage is minimized.
	
	\item Fast detection and reaction.
	\sysname is supposed to react on DDoS attack instantly after the DDoS attack starts.
	
	\item Automated reaction.
	The whole process of DDoS mitigation should minimize human interaction, that is, from DDoS detection to mitigation, the process should be automated without requiring human's decision making.
\end{enumerate}

\para{Assumptions}
The \sysname design is based on the following assumptions
that we argue are either reasonable or easy to be realized.
\begin{enumerate} [leftmargin=*, itemsep=1pt, parsep=2pt, topsep=4pt]
	\item \textit{Victim $S$ is best able to know its capacity to process incoming Interests under a specific prefix.}
	\noindent The server's capacity can be easily obtained based on the provisioned memory, CPU and other resources versus the time/space complexity of processing the Interest requests under the prefix.
	Such information collection has already been widely used in today's load balancing technologies.

	\item \textit{Victim $S$ is best able to know whether it is under a DDoS attack, the prefix that is under attack, and the types of attack Interest packets.}
	The victim server inherently has the most accurate judgments of a DDoS attack:
	By simply inspecting whether Interests under a prefix overwhelm the processing power, $S$ knows whether a prefix is under attack.
	Moreover, by processing an Interest packet, the server can immediately know \first whether the request is valid and \second whether the requested data is dynamic or static.

	\item \textit{An NDN router can be configured as a \sysname edge router.}

	In \sysname, an edge router is the gateway of an NDN enclave.
    This can easily be configured by the Internet Service Providers (ISPs) or the CDN providers at the network level (e.g. CPE, SOHO router, etc.) or by the device vendor (e.g., IoT vendor) at the overlaid NDN level.
%	The information can also be obtained through automatic means; for example, to learn whether it is connected to a client or not, the router can check the hop count of incoming packets from that interface.
\end{enumerate}

\subsection{\sysname Design Overview}

In a nutshell, \sysname reacts to feedback from a victim, traces back the specified traffic flows by checking NDN's forwarding state, and enforces traffic throttling at the edge.
The design is enabled by two major architectural properties of NDN.
\first The named traffic allows an expressive way for a victim to specify the attacking traffic and for \sysname to make fine-grained throttling.
\second The stateful forwarding provides run time insights into ongoing traffic flows so that \sysname can trace the traffic to precisely identify attackers.

\begin{figure}[t]
	\centering
	\includegraphics[width=0.48\textwidth]{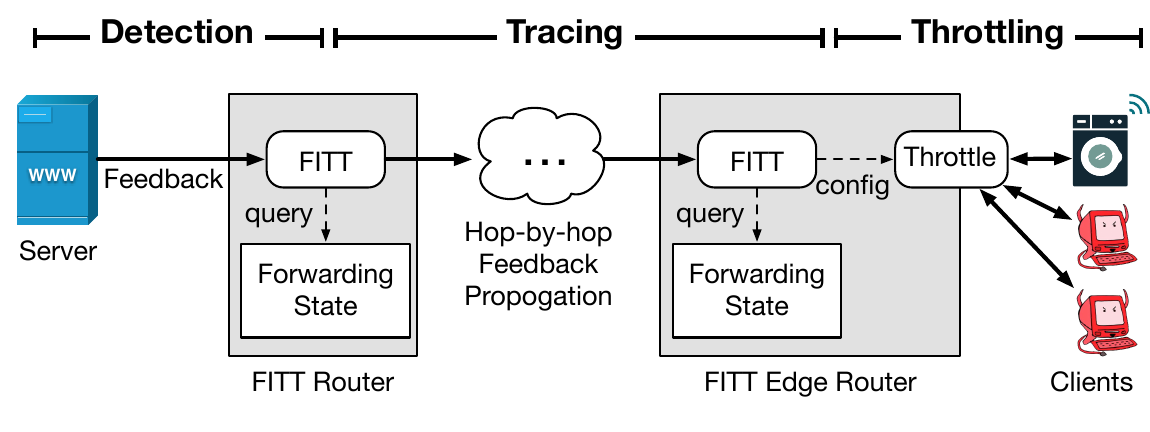}
	\caption{\sysname Overview}
	\label{fig:FITT-overview}
	\vspace{-3mm}
\end{figure}

\sysname works in three main steps as shown in Figure~\ref{fig:FITT-overview}.
We illustrate the three steps with the example topology in Figure~\ref{fig:toy},

\para{Step 1: Detection}
When server $S$'s service receives traffic more than the configured threshold, it sends out the feedback to its downstream router $R1$.
Once receiving the feedback, $R1$ parses the feedback and triggers the \sysname reaction based on the type of the attack.

\para{Step 2: Traceback}
In order to identify attacking sources, $R1$ checks its forwarding state.
Using names carried in the feedback from $S$, $R1$ can identify the attacking traffic flows by their names and then know the downstream routers ($R2$ and $R3$) from which the traffic is from.
$R1$ notifies the downstream routers and $R2$ and $R3$ will perform the similar procedures as $R1$ does.
In this way, \sysname reversely traces the attack traffic from $S$ all the way to \emph{edge routers} $R2$ and $R4$ where exact traffic senders are connected.

\para{Step 3: Throttling}
The edge routers will first notify these clients and then perform Interest throttling on suspect downstream interfaces within the specific prefix reported by $S$.
During the throttling, an edge router will check whether a client has changed its behavior or not (i.e. whether it lowers down its sending rate to the required value under the specified prefix).
The router can then relax or reinforce the limit, accordingly.

Multiple \sysname reactions can be triggered at the same time for different traffic prefixes (located on the same server or different servers) and different types of attacks.
On the edge, if traffic for a specific prefix from a suspect downstream interface is being throttled by multiple \sysname instances, a minimum allowed traffic value will be taken.

Since the mitigation is triggered by victim's feedback, \sysname can react immediately after the DDoS attack.
The latency for throttling to start is only a one-way trip time (0.5 RTT) from the victim server to clients.
In addition, since the whole process entirely operates over existing NDN forwarding plane, there is no man-in-the-loop for DDoS mitigation with \sysname.

\subsection{Explicit Feedback from the Victim}

A \sysname mitigation is triggered by a victim's feedback.
To be more specific, the feedback is carried by NDN NACK packet created by the victim server and sent downstream.
A feedback message carries the following information:
\begin{itemize} [leftmargin=*, itemsep=1pt, parsep=2pt, topsep=4pt]
\item \textbf{\emph{TYPE}}.
The \emph{type} code used to notify routers which type of Interest flooding is happening, i.e., invalid Interest attack, valid dynamic Interest flooding, or valid static Interest flooding.
\emph{TYPE} allows \sysname to react differently to different types of attacks.

\item \textbf{\emph{PREF}}.
This carries the \emph{prefix} under which the overwhelming traffic comes to the victim.
Based on \emph{PREF}, \sysname is able to push back attack traffic flows without affecting benign traffic.
Such a prefix should not be an arbitrary-length common prefix of the attack Interest packets; instead, the prefix should represent a unit of service or micro-service whose state of health (e.g., available compute capacity and network capacity) can be measured as a whole.
For example, if an application serves both request and dynamic requests, two sub name prefixes for request and dynamic requests should be used because server's capacity for handling these two types of requests is very different.

\item \textbf{\emph{RPS}}.
This is the receiving rate of valid Interests, i.e., \emph{request per second}, that the service can currently handle under the prefix \emph{PREF}.
\emph{RPS} can help to decide the strength of the traffic throttling in \sysname.
Its value can be obtained based on service's capacity versus the time/space complexity of the request.
\emph{RPS} is only sent when attack Interests are valid (Valid-S and Valid-D).

\item \textbf{\emph{InvalidNames}}.
\emph{Invalid Interest name list} is sent for invalid Interest attack only.
It contains exact invalid Interest names under the prefix \emph{PREF} that a server has received in recent past.
With \emph{InvalidNames}, routers can do exact match and identify attack packets in the forwarding state.
Optimizations can be applied to reduce the space complexity by sampling the list or utilizing Bloom Filters~\cite{bloomfilter} or regular expressions.

\end{itemize}

A \sysname mitigation will be triggered for each feedback message with a different $<$\emph{TYPE}, \emph{PREF}$>$.
A victim can send multiple feedback messages with the same $<$\emph{TYPE}, \emph{PREF}$>$ afterward to update the state, for example, the server can keep providing new \emph{InvalidNames} or update the value of \emph{RPS}.

Because of the importance of victims' feedback, the feedback messages should be authenticated.
There are several possible methods.
First, such authentication can be built up by pre-configure policies between CDN providers and service providers.
To be more specific, the feedback message can be signed with service provider's private key whose public key certificate is installed by the CDN nodes.
A second way is to compare the \emph{PREF} with the registered NDN route entry.
For example, in topology~\ref{fig:toy}, when router $R1$ receives a feedback message from the link connecting to $S$, reporting that the prefix \name{/com/target2} is under attack.
Assuming in $R1$'s routing table only prefix \name{/com/target} is registered on the interface towards $S$, $R1$ can drop the feedback message because the prefix \name{/com/target2} cannot be matched.

\subsection{Tracing Back to Exact Sources}

The router who receives a new feedback message will trigger the \sysname mitigation and trace back to exact attacking sources hop by hop.
In this process, the router will further propagate the feedback message with updated information towards downstream until reaching edge routers.
Importantly, all the procedures are performed to traffic under the prefix \emph{PREF} reported by the victim, so traffic under other prefixes will not be counted in.
In addition, for different types of attack, \sysname routers will operate differently.

\para{Invalid Interest Attack (Invalid)} When attack Interests are Invalid Interests, once receiving the feedback message, a router will first check \emph{InvalidNames} and find out the corresponding PIT entries from the forwarding state (i.e., Pending Interest Table).
Through these entries, the router learns the exact incoming interfaces and for each of these interfaces $i$, the router will generate a new fake Interest name list \emph{$\text{InvalidNames}_i$} which only contains invalid Interest names that were sent from interface $i$.
After that, the router will remove the invalid PIT entries and send a feedback message containing \emph{$\text{InvalidNames}_i$} to each interface $i$.

%The symbols used in this section are listed in Table~\ref{tab:notation}.
%\begin{table}[ht]
%	\vspace*{-0.2cm}
%	\caption{Notation Table}
%	\label{tab:notation}
%	\vspace*{-0.2cm}
%	\centering
%	\begin{tabular}	{ | m{2cm} | m{5.5cm}| }
%		\hline
%		$Num\_Matched$ & number of PIT entries that is under prefix $PREF$\\
%		\hline
%		$Faces_j$ & incoming interfaces recorded in the $j$th PIT entry\\
%		\hline
%		$Num\_Faces_j$ & number of incoming interfaces in the $j$th PIT entry\\
%	    \hline
%	\end{tabular}
%\end{table}

%The router will then calculate a weight $w_i$ for each incoming interface $i$ based on the number of incoming fake Interests through the interface.
%To calculate the weight $w_i$ for face i, we have
%\begin{equation}
%\label{equation:fake}
%w_i = \frac{1}{Num\_BL} \sum_{j=1}^{Num\_BL}\frac{1}{Num\_Faces_j} ~if~i \in Faces_j
%\end{equation}
%For each interface $i$, the router then sends a new NACK whose $T$ is  $w_i\times T$ and $BL$ is a pruned list that only contains fake Interests sent from interface $i$.

\para{Valid Interest Flooding (Valid-D or Valid-S)}
In valid Interest flooding, a router cannot directly distinguish the good traffic from the offending because they are all valid.
Consequently, the router will check all the traffic flows under the prefix \emph{PREF}, from which the router can get a set of suspect incoming interfaces.
The router will then calculate a weight $w_i$ for each suspect interface $i$ to distribute the \emph{RPS} to downstream routers behind these interfaces.
When the router has no statistical knowledge of how legitimate clients are distributed among these downstream routers, \sysname can simply adopt the simplest way of equally sharing the weight among all suspect interfaces and leave the further adjustment to edge routers -- the reinforcement throttling at the edge will help to further identify attacking traffic and amend the potential unfairness caused by the equal share.
\begin{equation}
\label{equation:valid}
w_i = \frac{1}{Num\_Suspect}
\end{equation}
where $Num\_Suspect$ represents the total number of suspect downstream interfaces.
When historical statistics of legitimate traffic flow's distribution among interfaces are available (e.g., maintained by the CDN service providers), the weight can be computed in smarter ways.
After that, for each suspect interface $i$, the router will send a new feedback message containing a weighted \emph{$\text{RPS}_i$} $ = w_i \times \text{\emph{RPS}}$.

%In this way, all the routers along the Interest sending path recursively receive and generate new NACKs that will be propagated to further downstream routers.
%Finally, the DDoS report that originated from the server will arrive at all \sysname edge routers.
%
%\begin{figure}[ht]
%	\centering
%	\includegraphics[width=0.4\textwidth]{figures/flow}
%	\caption{Router's Process Logic After Receiving a NACK}
%	\label{fig:flow}
%\end{figure}

\subsection{Fine-grained, Reinforcement Throttling at the Edge}

In our design, only \sysname edge routers play the role of rate limiting.
This is because, on one hand, we cannot trust a client's device to take actions - it could be compromised as well.
On the other hand, an upstream router should not perform rate limiting for the following reasons:
\begin{itemize} [leftmargin=*, itemsep=1pt, parsep=2pt, topsep=4pt]
	\item When the traffic volume under a target prefix increases, upstream routers do not have enough knowledge to tell whether it is because of the misbehaving downstream routers or new clients have joined.
	\item When legitimate clients are behind a downstream router, compared with edge routers, upstream router actions will also hurt legitimate clients.
\end{itemize}
Note that collateral damage is one important reason for DDoS remediation attempts in IP network (using FlowSpec, RTBH, etc.) to suffer and be maligned.

When a \sysname edge router receives a feedback message from the upstream, it will perform traffic throttling to suspicious downstream interfaces (connecting to the client end points) where attack traffic is from and monitor traffic from these interfaces for further adjustment.
To be specific, the edge router will first calculate a permitted sending rate \emph{$\text{RPS}_i$} for the traffic under prefix \emph{PREF} from each suspicious interface $i$.
\begin{center}
	\vspace{-0.5cm}
	\[
	Limit_i =
	\begin{dcases}
	\text{\emph{RPS}}_i,& \text{if \emph{TYPE} = VALID-S or VALID-D}\\
	0,& \text{if \emph{TYPE} = INVALID}
	\end{dcases}
	\]
\end{center}
In the valid Interest flooding, the calculation of \emph{$\text{RPS}_i$} at the edge is the same as in upstream \sysname routers.
In contrast, since invalid Interests are almost only used by attackers, the edge router will drop all traffic flows under \emph{PREF} from the suspicious interface.

\para{Reinforcement Throttling}
We believe that the legitimate clients are willing to obey the DDoS control and lower their sending rate of Interests accordingly while attackers may not abide by this.
Therefore, in \sysname, the router will first sends a feedback message to each suspect interface as a notification and then starts throttling traffic by dropping Interest packets in random to ensure:
\begin{center}
	$\forall i \in range(1, n) ~~ \text{\emph{RPS}}_{(i,~P)} \leq Limit_i $ \\
\end{center}
where $RPS_{(i,~p)}$ is the Interest sending rate from interface $i$ under prefix \emph{PREF} and $Limit_i$ is the permitted Interest sending rate of interface $i$.

Once receiving a feedback message from the gateway router, legitimate clients are supposed to comply by lowering down their Interest sending rate under the prefix \emph{PREF}, while the bots may not obey the rules, prompting the router to perform reinforcement throttling.
In case of valid Interest (Valid-S, Valid-D) flooding, the router will monitor the sending rate of each suspicious interface and perform the reinforcement throttling:
\begin{itemize} [leftmargin=*, itemsep=1pt, parsep=2pt, topsep=4pt]
	\item If a sender lowers its Interest sending rate $\text{\emph{RPS}}_{(i,~p)}$ to comply with $Limit_i$, the router will remove the throttling over this client in the next time period of throttling.
	\item If a sender does not comply, the router will reset the limit to $\frac{1}{2} \times Limit_i$.
\end{itemize}
The adjustment will help \sysname to further restrict the attacking traffic and relax the limit on legitimate traffic by adjusting the throttling to be fair.
As shown in simulation results in Section~\ref{sec:evaluation:valid}, the reinforcement will quickly block all attacking traffic and let legitimate traffic recover from the throttling

It is possible that bots may use intelligence to analyze and attempt to circumvent the throttling, but \sysname already succeeds if the bots cannot increase Interest sending rate, thus greatly reducing damage.
Essentially, \sysname forces bad entities to comply.
In the evaluation section~\ref{sec:evaluation:smart}, we simulate ``smart" attacker and the results confirm the above statement.

\para{Fine-grained Throttling}
The fundamental benefit of the fine granularity of \sysname is twofold:
\begin{itemize} [leftmargin=*, itemsep=1pt, parsep=2pt, topsep=4pt]
	\item By explicitly setting \emph{PREF} in a feedback message, \sysname squelches the traffic under the prefix \emph{PREF} only.
	All the other services provided by the victim server will not be affected.
	\item \sysname throttles traffic sent by a suspect client to the prefix \emph{PREF}, only, letting the clients communicate to other services.
\end{itemize}

Consider the example in Figure~\ref{fig:toy}.
When service \name{/com/target/service1} is under attack, \sysname will limit the traffic to this service to the expected volume \emph{RPS} as configured by the victim $S$, or block the traffic consisting of invalid Interests.
In the throttling, non-attacking clients like $C4$ and $C5$ can use $S$'s services (e.g., \name{/com/target/service2}) that are not attacked as normal, reducing the collateral damage on $S$.
Assume $C3$ is a compromised smart home device, e.g., a home camera.
Though $C3$ is compromised to send attack traffic to $S$, \sysname is able to stop its DDoS traffic and at the same time, does not bother its normal functions, e.g., it can still upload the surveillance video records to the smart home controller.

\section{Leap-Forward Deployment}\label{sec:cdn}

\begin{figure*}[ht!]
	\centering
	\includegraphics[width=5.2in]{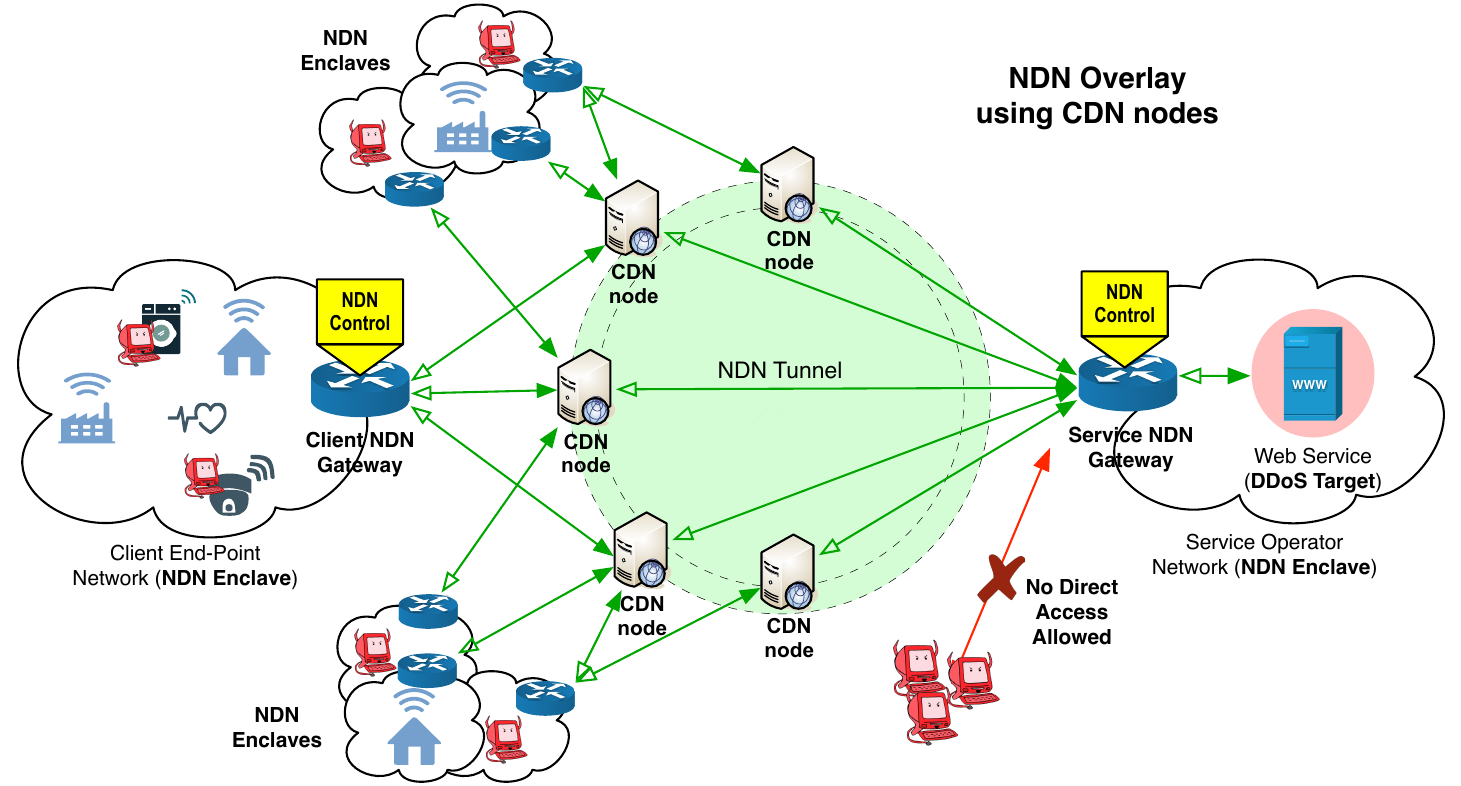}
	\caption{Incremental (leap-forward) deployment of \sysname using a CDN as an overlay to pass traffic between NDN enclaves. Note that attackers are not able to directly reach the target because the edge NDN gateways communicate only with the CDN overlay nodes.}
	\label{fig:Leap_forward_arch}
	\vspace{-5mm}
\end{figure*}

%\zhiyi{Is it possible to split this long para into several paras for the sake of reading?}
%\zhiyi{I make some ordering and splitting paragraphs based on what I understood. Please double check.}

While in a perfect world, we would be able to immediately deploy NDN and \sysname to safeguard our networks from the menace of DDoS, it is more realistic to assume that an incremental deployment of NDN starting from the edges and slowly pushing towards the Internet core is more pragmatic. 
NDN and \sysname present an option for immediate incremental deployment, which elevates many of their advantages to near-term objectives. In this paper, we observe that the suitability of NDN's architecture to perform DDoS remediation is not just a parallel benefit to its suitability to performing CDN functions; rather, we posit that existing CDN deployments are opportune infrastructure to enable broad deployment of NDN. Importantly, many CDNs' existing roles as MaaS providers suggest the potential alignment of costs with incentives to performing both CDN and MaaS. 
\begin{itemize} [leftmargin=*, itemsep=1pt, parsep=2pt, topsep=4pt]
	
	\item First, an upgrade of a CDN to support NDN could provide a synergistic benefit to Internet services that want protection from DDoS, and to the CDN/MaaS provider. 
	Multiple aspects of the synergy that exists between DDoS mitigation and NDN.
	% should serve as an incentive for CDN/MaaS providers to deploy NDN in their infrastructure. 
	For example, one of the devastating aspects of volumetric DDoS attacks occurs when attack traffic is backhauled from distributed sources towards destinations (whether the destinations are victim services, or even to MaaS scrubbing centers)~\cite{osterweil202021}.  
	Deployment of NDN across a CDN/MaaS would let that provider shed attack traffic at the edges before it starts to aggregate across transit links. 
	
	\item Second, CDN/MaaS providers already shoulder the computation and network requirements that NDN would require. For the TCP/IP Internet, CDN/MaaS providers have already been terminating TLS connections and proxying connections from clients to service infrastructures. 
	An NDN upgrade would fit the operational footprint that large CDNs already have, would not necessitate deployment of additional resources, but would also fundamentally enhance DDoS service offerings. 
	
	\item Furthermore, among the properties of NDN's information centric architecture is its inherent capability to cache data near its consumers, architecturally. This fundamental advantage is poised to not only be a pivotal feature, but also a deployment incentive for large Content Delivery Networks (CDNs) providers. In today's TCP/IP Internet, CDNs are vast networks and deployments that operate as overlay services for end-users. CDNs exist in TCP/IP networking (above the architectural layer) to efficiently deliver content to users today. CDN services typically involve large network infrastructures and deployments. They often perform caching of content to locations that are geographically distributed, as well as geographical load-balancing of requests from clients so that network latency can be minimized. In today's Internet CDNs exist as overlay technologies, and some have proposed that NDN is well suited to implement these functions in the network architecture, itself~\cite{cao2016fetching,jiang2014ncdn,ma2014tentative}. Furthermore, many operational CDNs, today, also perform DDoS mitigation offering commercial DDoS/MaaS services: Akamai~\cite{akamai-prolexic}, Cloudflare~\cite{cloudflare-ddos}, and Neustar~\cite{neustar-ddos} to name a few.

\end{itemize}

The resulting NDN network would have global scope and in-network caching in the topologically distributed regions. 
In general, incremental deployment models become more realistic when they align their costs with incentives. That is, those who deploy new mechanisms are more likely to do so when they anticipate direct benefits from doing so. By contrast, deployments like ingress and egress filtering (BCP-38~\cite{RFC2827} and BCP-84~\cite{bcp84}) illustrate slow adoption, arguably, because those deploying them do not gain any direct benefits. Conversely, service providers already expend resources and money to combat DDoS by either provisioning large amounts of excess bandwidth or by contracting with commercial DDoS mitigation providers~\cite{akamai-prolexic, neustar-ddos, cloudflare-ddos}. Service providers whose applications may already be in a position to benefit from migrating to NDN's architecture would gain \emph{additional} benefits by deploying NDN with \sysname and thereby being able to shed large amounts of DDoS traffic.

Therefore, we propose an incremental deployment (see Figure~\ref{fig:Leap_forward_arch}) that leverages CDN to connect NDN routers located at the edges of the network. The two ends speaking NDN are enough for \sysname to effectuate the DDoS mitigation. This means that our approach requires NDN gateways on both sides of the CDN overlay: the client end-points and the service provide/operator side. On the client end-point, the NDN gateway can both forward traffic from existing NDN devices and translate at the application level using an application proxy traffic from traditional TCP/IP devices. Such a translation is akin to having an application proxy (i.e. web proxy) and it would allow the client NDN gateway to perform enforcement without having to worry about packet classification or deep packet inspection. On the service operator side, the NDN gateway can serve traffic directly to Web Service. What's more, service operators who migrate their services to NDN bolster each others' NDN deployments, as those clients independently augment each others' deployments (through facilities like shared caching and shared routing infrastructure). In particular, we observe that serendipitous IoT deployments of NDN, which may already be underway, could benefit other services whose providers have (or will) independently embraced NDN for this reason. That is, an NDN-enabled service may shed DDoS traffic from would-be attack nodes that might otherwise be bots in Mirai. By enabling NDN at the edges in home routers and IoT deployments would place the \sysname mitigation machinery very near to some of the Internet's most voluminous DDoS sources for all NDN applications (not just IoT). We believe that it is demonstrably feasible for independent service operators to overcome network protocol ossification and migrate (at least portions) of their production traffic to NDN. Furthermore, we show later in this section that the \sysname/NDN deployment outperforms existing CDN caching schemes because it is performed as a network function deeper into the stack. Thus, by enabling NDN at the edges, we get both performance and security benefits without sacrificing functionality.

The deployment of NDN/FITT may also not need to bother the change of existing end-point applications running in NDN enclaves. To be more specific, the gateway of NDN enclaves can be an NDN forward/reverse proxy. For clients, e.g., IoT devices, to talk to a remote service, the client's gateway server plays the role as a forward proxy which turns application-layer request (e.g., HTTP GET request) into an NDN Interest packet and sends it out to the server over the NDN tunnel; while for the service provider side, the server gateway router serves as a reverse proxy parsing NDN Interest packets back to normal request. When servers send back the response, the process is similar. Given the commonalities between NDN's Interest-Data exchange and today's widely-used request-response model in application layer protocols (e.g., HTTP, RPC), such proxy and reverse proxy are deployable and the cost can be reasonable because it does not require any hardware change -- all NDN proxy deployment is software installation in user space.

\subsection{Implementation of \sysname Prototype}

\sysname has been implemented in C++ as a plug-and-play module in current NDN Forwarding Daemon (NFD).
To be more specific, the core logic of \sysname is designed to be running on each NDN forwarder as a forwarding strategy.
In an NDN forwarding module, forwarding strategies decide the forwarding operations.
Importantly, adding a new forwarding strategy requires no modification to the forwarding module design and the new strategy can be turned on with run-time configurations.

The feedback message is realized with NDN NACK packets because NACK is already being used for hop-by-hop feedback (e.g., reporting errors in forwarding like no route).
Extending NACK's syntax for \sysname does not require any change of existing NDN forwarding logic.

\para{State  in \sysname}
To implement \sysname, we modified the NFD to record an additional internal state, \sysname record, on receiving a feedback message with a new $<$\emph{TYPE}, \emph{PREF}$>$. 
A \sysname record records \emph{TYPE} (1 Byte),  \emph{PREF} (variable length), and \emph{RPS} (4 Bytes) from the feedback message and a timestamp (4 Bytes) used for expiration.
Therefore, a record usually takes less space than a pending Interest entry.
A \sysname record table is maintained when there are multiple different ongoing \sysname mitigation instances.
The primary key of the table is $<$\emph{TYPE}, \emph{PREF}$>$ and the space complexity is $O(n)$ where n is the number of current \sysname instances.
The record will be removed by the router for some time after receiving the original feedback message unless new feedback messages with the same $<$\emph{TYPE}, \emph{PREF}$>$ arrive.
The time period is decided by RevertTimer as discussed in the rest of the section.

\para{Timers in \sysname}
\label{sec:design:timer}
\sysname utilizes two types of timers, \emph{RevertTimer} and \emph{RateLimitTimer}.
These timers affect the overhead of \sysname but will not affect the outcome of \sysname mitigation.
These timers can be configured by the DDoS mitigation service provider according to their provisioned resources.
\begin{itemize} [leftmargin=*, itemsep=1pt, parsep=2pt, topsep=4pt]
\item A \emph{RevertTimer} decides how long a router should keep the \sysname records.
The timer is set for each creation and update of the \sysname records.
On receiving a new feedback message, the router checks whether there is an existing RevertTimer for the \sysname mitigation with the same $<$\emph{TYPE}, \emph{PREF}$>$.
If yes, the router will update the timer instead of creating a new one.
When the timer fires alarms, the record will be removed from the router.
%\sichen{I'm not sure of the effect when this timer expires. Will there still be reaction on existing DDoS traffic/Clients in black list. Also, if this is a expiration timer of FITT records, do we need one for each FITT record?}
% as discussed in Section~\ref{sec:dis:adjacentnack}.

\item A \emph{RateLimitTimer} is maintained by edge routers only.
It decides the time interval of statistics on Interest sending rates from each downstream interface.
After the RateLimitTimer expires, the gateway router will remove the limit of ``good'' clients and strengthen the limit of bad ones.
This timer is periodically reset until all clients either behave well or be totally blocked for the prefix \emph{PREF}.
\end{itemize}

\begin{figure}[t]
	\centering
	\includegraphics[width=0.48\textwidth]{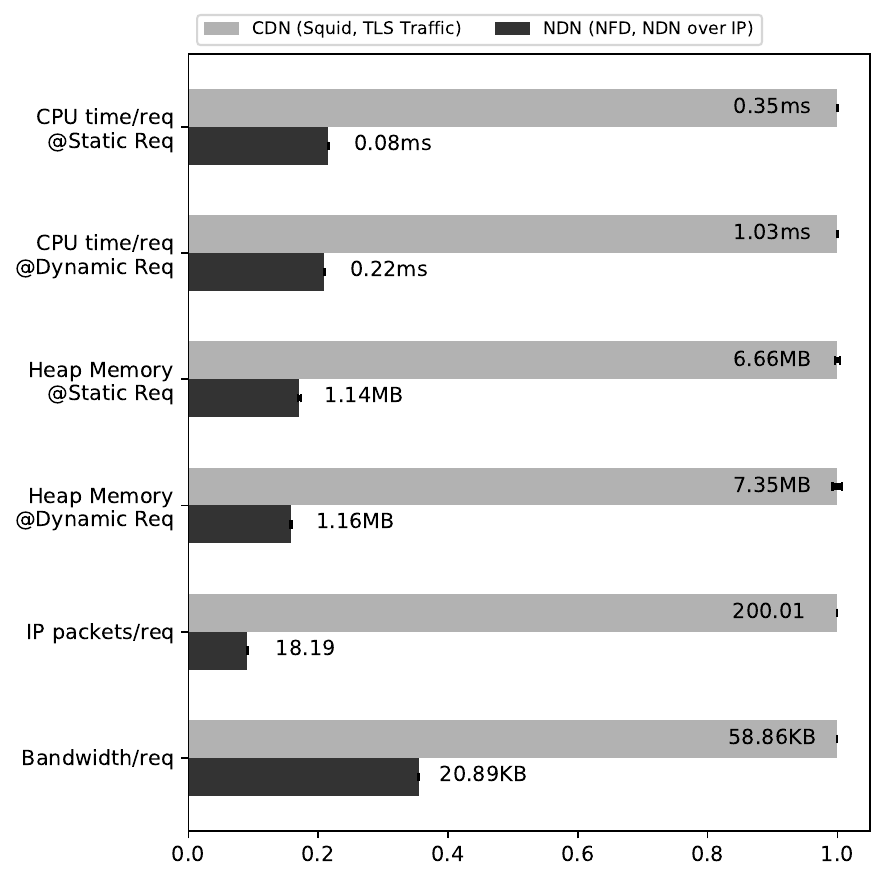}
	{\footnotesize \\
			\first Heap memory consumption for static and dynamic requests (static request will be satisfied by both CDN cache and NDN cache while dynamic request will be forwarded by CDN/NDN proxy).\\
			\second CPU consumption for static and dynamic requests. \\
			\third Number of packets and traffic received (TLS over traditional TCP/IP versus FITT tunneling NDN packets from an NDN enclave)
		\par}
	\caption{Comparing a vanilla CDN proxy running Squid and the same proxy running a prototype of \sysname on top of NDN}
	\label{fig:cdn_ndn}
	\vspace{-5mm}
\end{figure}

\subsection{Performance Comparison of \sysname vs CDNs} \label{sec:cdn_ndn_eval}

We simulated a topology similar to Figure~\ref{fig:Leap_forward_arch} where a CDN infrastructure is used for static and dynamic page delivery for web services. In the pure CDN scenario, the CDN proxies are running the latest version of Squid~\cite{Squid} and for our approach the same CDN proxies run a version of our \sysname prototype on top of NDN. Our aim was to be able to evaluated the computation and communication overhead of NDN forwarding daemon (NFD), an open-source NDN network forwarder, and Squid, one of the most widely used web proxy systems. Specifically, we performed multiple simulations using today's practice of MaaS and NDN/FITT's DDoS mitigation with the same hardware settings. In our experiments, we used computers equipped with Intel Core i9 4.6GHz processor with 32GB DDR4 RAM. We also used the same number of clients requesting the same amount of data and computation: six (6) clients requesting for both static content and dynamically-computed content of 2KB at the same rate simultaneously. 

As shown in Figure~\ref{fig:cdn_ndn}, we first present the CPU and memory use of a single-thread Squid and NFD/FITT under the same load. The plot indicates that the combination of NFD with FITT has an advantage in terms of resource consumption when compared to Squid: under the same load of traffic, Squid consumes an estimated of about five times (5x) more memory than \sysname. We observed a similar trend when we measured the computation footprint on each of the CDN proxies: \sysname requires a mere 20\% of the processing time when compared to Squid. In addition, we observed the total amount of traffic at the networking layer (IP) involved assuming a CDN proxy running Squid receiving TLS traffic vs NDN traffic (similar to Figure~\ref{fig:Leap_forward_arch}. 
In both cases we made the assumption that traffic is forwarded over a single hop between the CDN and the MaaS scrubbing center. This is the worst case scenario for us because, in practice, MaaS scrubbing centers can be deeper into the network in a centralized location and several network hops away from the edge CDN proxies. Even under that assumption, our experiments indicate that in order to fetch the same amount of content and computation results, Squid over TLS requires around ten times (10x) the amount of packets resulting in an almost three times (2.8x) bandwidth overhead compared with NFD/\sysname over IP-overlaid NDN. As we mentioned, since MaaS service requires TLS-terminating traffic forwarding on both the CDN proxies and MaaS scrubbing centers, the computation and communication overhead can be much higher than what we report here for the vanilla CDN implementation making \sysname a much more desirable option.
    %\sichen{shall we use "MaaS scrubbing centers"}

\subsection{Towards Full Deployment: Overcoming Ossification}

A common lament on the Internet has been that network protocols evolve very slowly, or tend to be ossified~\cite{1432642, permanent2019Sigcomm}. Recent deployment successes in other network protocols~\cite{quic-sigcomm17} has illustrated that this impasse can be overcome by providers who control (i.e. implement and deploy)
both ends of a service (the client and server sides).  When implementing mobile apps, the provider has the ability to choose both ends of the network protocol. While applications that depend solely on web browsers must often remain backward-compatible with TCP/IP, mobile clients can often implement service-specific code. The deployment of Google's QUIC~\cite{quic-sigcomm17} provides a timely example of this flexibility.
In that case, deployment grew quickly with Google's ownership of the transactions.\footnote{Google was able to implement QUIC on its mobile platforms and its Chrome browser, but maintained TCP/IP support for other browsers.} Legacy TCP support was maintained, but QUIC was treated as preferred where QUIC-compatible clients were used.
We observe that this tactic is equally available to NDN, through mobile applications.
As a migration path, and to maintain backward compatibility, service providers could bifurcate their deployments and offer TCP/IP services on separate infrastructure.
Then, under cases like large DDoS attacks, TCP/IP could be serviced by different infrastructure, and all NDN/\sysname infrastructure could remain unencumbered by attack traffic,
while TCP/IP remediations are enacted on the legacy infrastructure.

%\begin{figure}[t!]
%	\centering
%	\includegraphics[width=0.38\textwidth]{simulation/plot-cpu-bar}
%	\caption{Processing times for static (cache hit) and dynamic (cache miss) operations comparing a vanilla CDN proxy running Squid and the same proxy running a prototype of \sysname.}
%	\label{fig:cdn_ndn_cpu}
%%	\vspace{-5mm}
%\end{figure}
%
%
%
%\begin{figure}[t!]
%	\centering
%	\includegraphics[width=0.38\textwidth]{simulation/plot-traffic}
%	\caption{Number of packets and traffic generated when comparing web traffic over TLS using a vanilla CDN proxy running Squid utilizing traditional TCP/IP versus the same proxy running a prototype of \sysname tunneling NDN packets towards an NDN enclave. In both cases we utilized web requests to the same web pages from a client connecting to CDN.}
%	\label{fig:cdn_ndn_traffic}
%	\vspace{-5mm}
%\end{figure}

%\input{sections/rational}
%\input{sections/deploy}
\section{Evaluation of \sysname \& NDN's DDoS Resilience}
\label{sec:evaluation}

\begin{figure} [t]
	\centering
	\captionsetup{justification=centering}
	\includegraphics[width=0.4\textwidth]{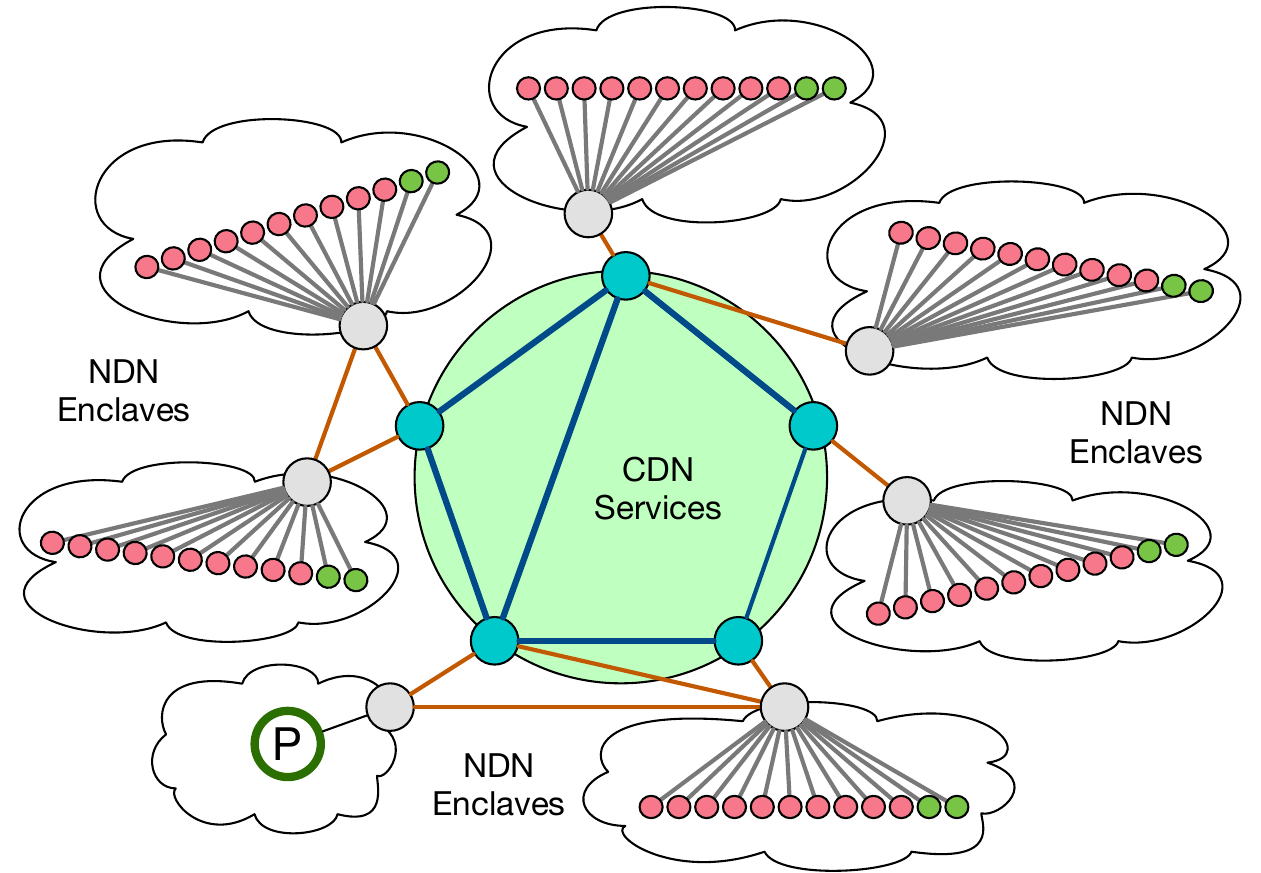}
	\caption{Incremental deployment topology for simulations}
	\label{fig:topology}
	\vspace{-0.5cm}
\end{figure}

Besides the prototype implementation over the latest stable version of NFD, we also implement \sysname in C++ over ndnSIM~\cite{mastorakis2017ndnsim}, which is a NDN simulation platform based on NS-3.
We first demonstrate NDN's DDoS resilience to valid static Interest flooding and then evaluate \sysname under different types of attacks.
The simulation results show that after the DDoS starts, \sysname can effectively control the traffic to the victim as expected within seconds (less than 2 seconds under our simulation settings), and ensure that over 99\% of the attack target(s) incoming traffic is from legitimate clients after a short period of time.

In addition, we simulate the scenarios when \first multiple DDoS mitigation instances are happening at the same time, \second different traffic flows are presented and only one of them is throttled, and \third attackers are rogue and follow the DDoS control of \sysname.
The results shows \sysname can perform fine-grained DDoS mitigation and handle comprehensive attack scenarios.

\para{Simulation Topology}
We simulate the incremental deployment of NDN/\sysname using a CDN as an overlay between NDN enclaves.
As shown in Figure~\ref{fig:topology}, the blue nodes are CDN nodes that are aware of NDN and FITT, the gray nodes are gateway routers of NDN enclaves.
Behind each gateway router, we simulate 10 compromised IoT devices and 2 honest IoT devices.
Since \sysname follows a divide-and-conquer strategy in traffic throttling, the scale of the network topology does not affect evaluation results of \sysname much.
We make the service globally reachable, which means all users have means to learn the name and express Interest packets towards the service.
For sake of simplicity, we use the prefix $P$ to represent the service in rest of the section.

\para{Simulation Result Notation}
In the simulation result plots, we use the red dashed line to represent attackers' sending rate (RPS), the blue solid line to represent DDoS target's receiving traffic rate (RPS), the green dot-dashed line to represent legitimate clients' sending rate (RPS).
Therefore, when blue solid line meets the green dot-dashed line, all traffic arriving at the DDoS target is from legitimate clients.
The area below the red dashed line and above the blue solid line represent the DDoS mitigation provided by \sysname over NDN.

\subsection{NDN: Resilience to Valid-S Interest Flooding}
\label{sec:evaluation:agg-cache}

\begin{figure} [t]
	\centering
	\begin{subfigure}{.23\textwidth}
		\centering
		\includegraphics[height=4cm]{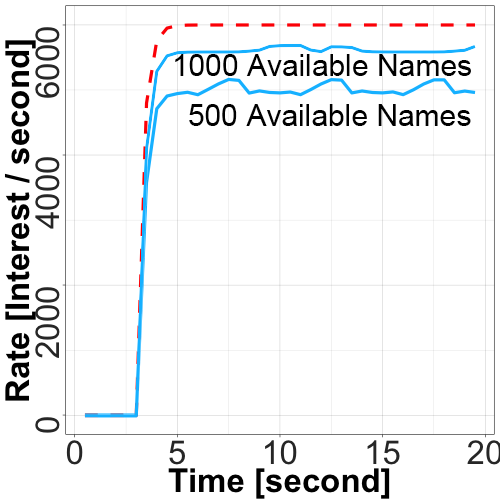}
		\caption{Interest Aggregation}
		\label{fig:no-cache}
	\end{subfigure} %
	~
	\begin{subfigure}{.23\textwidth}
		\centering
		\includegraphics[height=4cm]{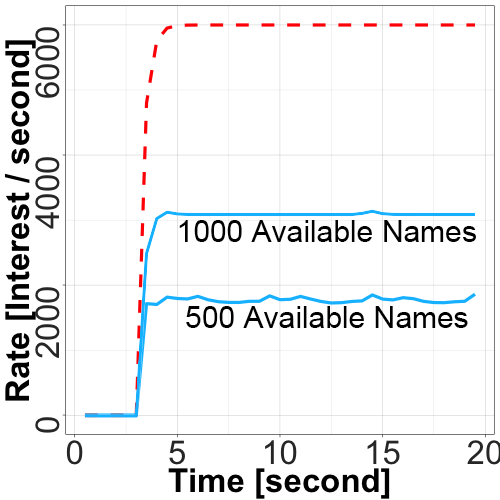}
		\caption{In-network Cache}
		\label{fig:with-cache}
	\end{subfigure}
	\vspace{2pt}
	{\footnotesize \\
		\textbf{Simulation Settings:}
		NDN deployment without \sysname.
		Each attacker (60 in total) start sending Valid-S attacking Interests at 100 pkt/s from second 3 with available number of data names 500 and 1000, respectively.\par}
	\caption{NDN's DDoS Resilience to Valid-S Interest Attack}
	\vspace{-0.3cm}
	\label{fig:interest-aggregation-caching}
\end{figure}

Figure~\ref{fig:interest-aggregation-caching} demonstrates NDN's DDoS resilience to static (Valid-S) Interest flooding with NDN's intrinsic properties, i.e., Interest aggregation and in-network cache.
To be more specific, we first disabled cache in all routers so that the result will only be affected by NDN's Interest aggregation.
As shown in Figure~\ref{fig:no-cache}, NDN can withhold traffic from attackers (red dotted line) to the server (blue solid line).
The more available names are used, the less traffic NDN can retain.
We then introduce cache capacity of 200 data packets in Figure~\ref{fig:with-cache}.
It is apparent that the number of Interests reaching $P$ decreases because of the caching capacity, which is because intermediate nodes along the path will serve future same Interests with cached Data (the freshness of cached Data is 4 seconds in our simulation).

The two figures indicates that the effect of Interest aggregation and cache is lower when an attacker can use a bigger set of Interest names to attack the victim.
This is because larger the name set, smaller the chance of two Interests carrying the same name and smaller the chance to hit a previous cached Data packets.

\subsection{\sysname: Invalid Interest Attack}

\begin{figure} [t]
	\centering
	\begin{subfigure}{.23\textwidth}
		\centering
		\includegraphics[height=4cm]{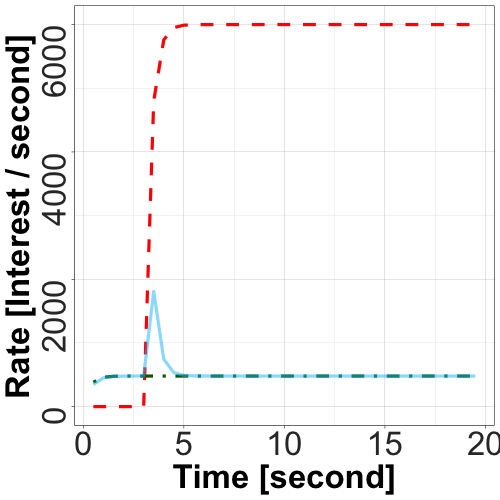}
		\caption{Invalid Interest DDoS}
		\label{fig:fake-attack}
	\end{subfigure} %
	~
	\begin{subfigure}{.23\textwidth}
		\centering
		\includegraphics[height=4cm]{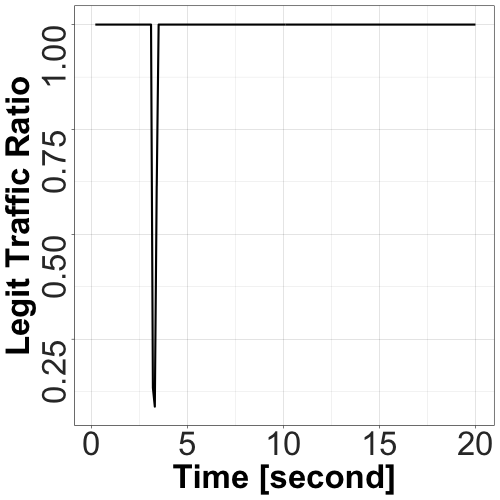}
		\caption{Legitimate Traffic Ratio}
		\label{fig:fake-percentage}
	\end{subfigure}
	\vspace{2pt}
	{\footnotesize \\
	\textbf{Simulation Settings:} NDN deployment with \sysname.
	Each attacker (60 in total) starts sending invalid Interests at 100 pkt/s from second 3. Legitimate clients' sending rate is 40 pkt/s starting from second 0.\par}
	\caption{\sysname mitigation of invalid Interest attack}
	\vspace{-0.3cm}
	\label{fig:fake-ddos}
\end{figure}

We first study \sysname's performance against Invalid Interest (Invalid) DDoS attack.
As shown in Figure~\ref{fig:fake-attack}, initially, $P$ only receives Interests from legitimate clients (green dot-dashed line).
After 3 seconds, attackers start the DDoS by sending Invalid Interests (the red dashed line) to $P$.
As depicted by the plot, $P$'s incoming traffic line (blue solid line) immediately goes up after the attack but soon goes down and merge the legitimate clients' outgoing traffic line (green dotted line).
Therefore, \sysname can eliminate the invalid Interest DDoS traffic in a short time.
The effectiveness is because in invalid Interest DDoS attacks, \sysname can accurately identify attackers by \emph{InvalidNames} carried in feedback messages and throttle their attack traffic.
Figure~\ref{fig:fake-percentage} shows that after \sysname reaction, all the traffic received by $P$ is from legitimate clients.

\subsection{\sysname: Valid Interest Flooding}
\label{sec:evaluation:valid}

\begin{figure} [ht]
	\centering
	\begin{subfigure}{.23\textwidth}
		\centering
		\includegraphics[height=4cm]{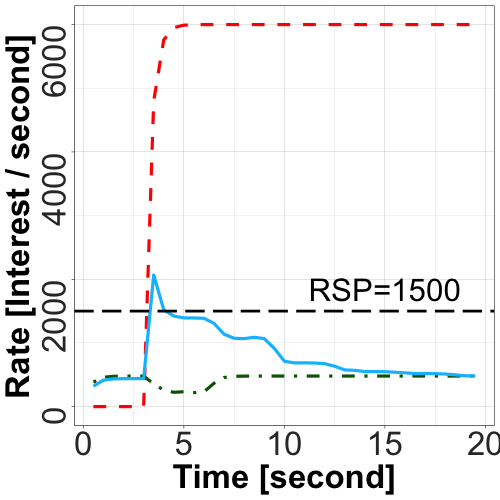}
		\caption{Valid-D Interest DDoS}
		\label{fig:valid-attack}
	\end{subfigure} %
	~
	\begin{subfigure}{.23\textwidth}
		\centering
		\includegraphics[height=4cm]{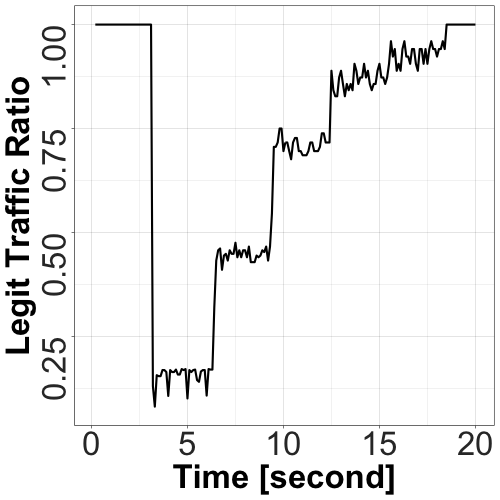}
		\caption{Legitimate Traffic Ratio}
		\label{fig:valid-percentage}
	\end{subfigure}
	\vspace{2pt}
	{\footnotesize \\
		\textbf{Simulation Settings:}
		NDN deployment with \sysname.
		Each attacker (60 in total) starts sending Valid-D Interests at 100 pkt/s from second 3. Legitimate clients' sending rate is 40 pkt/s starting from second 0.
		We let $P$'s capacity be 1.5K RPS and \emph{RateLimitTimer} be 3 seconds.
		\par}
	\caption{\sysname mitigation of Valid Interest flooding}
		\label{fig:valid-ddos}
		\vspace{-0.3cm}
\end{figure}

We then simulate valid Interest flooding.
To remove the effect of in-network cache, we disabled all router's cache.
As such, Valid-S and Valid-D Interest flooding become the same because all of them will arrive at the DDoS target $P$.
The results of Valid-D Interest flooding are shown in Figure~\ref{fig:valid-ddos}.

\begin{figure*} [ht]
	\centering
	\begin{subfigure}{.28\textwidth}
		\centering
		\includegraphics[height=4cm]{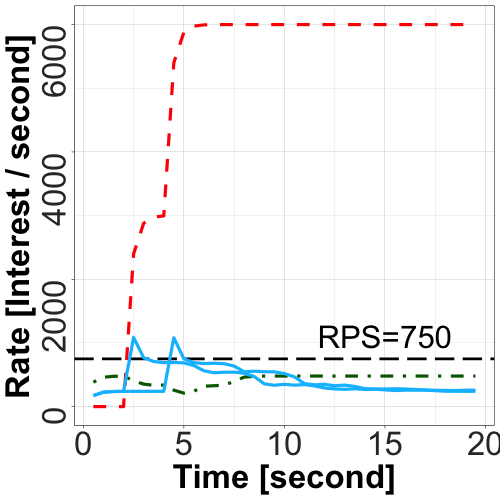}
		\caption{Two-victim DDoS attack}
		\label{fig:two-valid-attack}
	\end{subfigure} %
	~
	\begin{subfigure}{.28\textwidth}
		\centering
		\includegraphics[height=4cm]{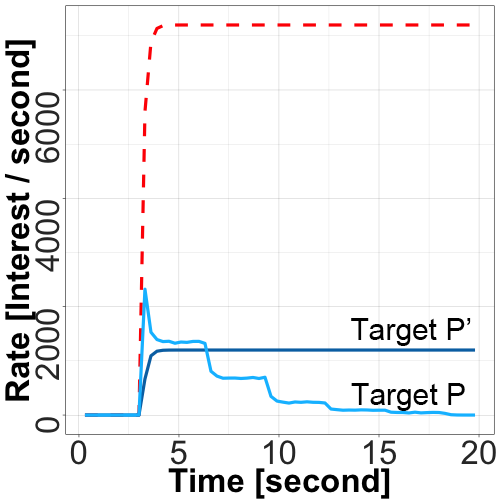}
		\caption{Fine-grained throttling}
		\label{fig:granularity-ddos}
	\end{subfigure}
	~
	\begin{subfigure}{.4\textwidth}
		\centering
		\includegraphics[height=4cm]{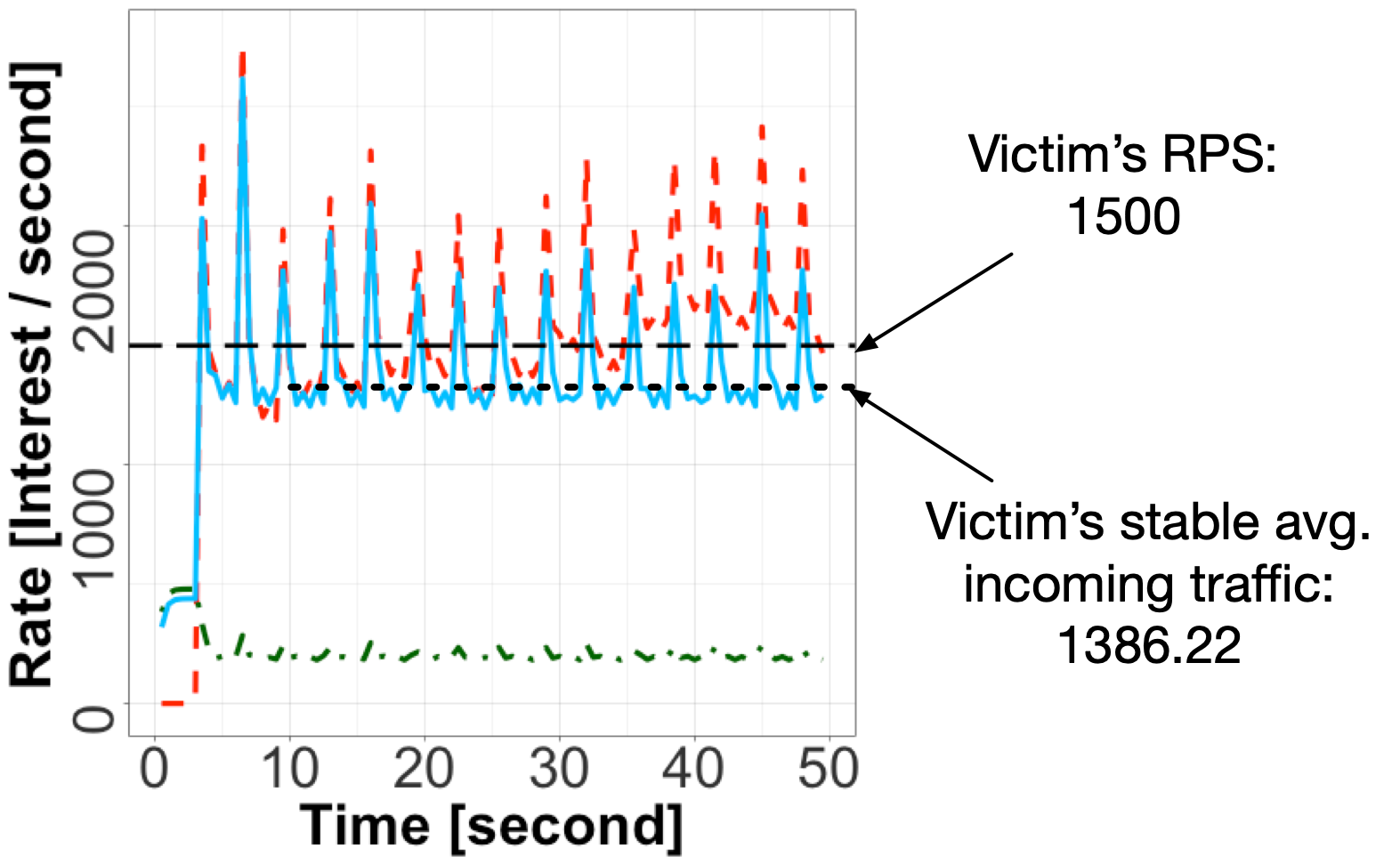}
		\caption{``Smart" attackers}
		\label{fig:smart-attacker}
	\end{subfigure}
	\vspace{-0.2cm}
	\caption{\sysname mitigation under different scenarios}
	\vspace{-0.5cm}
	\label{fig:two-valid-ddos}
\end{figure*}

Since the router cannot tell good traffic from bad traffic when DDoS starts, legitimate clients are also limited.
After receiving the \sysname feedback messages, legitimate clients will abide by the control placed and lower down their sending rate until the router determines them to be legitimate and free the limits, explaining why the green dot-dashed line goes down in the first several seconds of the attack and then back to the normal later.
As for attackers, as shown, the traffic received by the victim drops periodically (every 3 seconds), which confirms the \sysname's reinforcement throttling:
\sysname will halve the limit on attackers until all the attackers' traffic to the reported prefix are totally blocked.
At the end of the mitigation, $>$99\% of the Interests received by the victim are from legitimate clients (Figure~\ref{fig:valid-percentage}).

\subsection{\sysname: Multiple Attacks to Different Prefixes}
\label{sec:evaluation:multiple}

\sysname is designed not only to handle single DDoS attack but also comprehensive DDoS attacks, i.e., attacks to different prefixes, starting at different time and using different types of Interests.
In this simulation, we evaluate a more complicated attack scenario where half of the attackers attack service $P$ with Valid-D Interests starting from second 2 and another half attack the another service $P'$ with Valid-S starting from second 4.
We found the simulation results of scenarios when $P$ and $P'$ are located on the same node or different nodes are almost the same.
Figure~\ref{fig:two-valid-attack} shows the result when $P$ and $P'$ are running on the same server.

As shown, when multiple \sysname mitigation instances take place, \sysname can effectively control the DDoS traffic from both attacks at the same time.
For each victim server, the incoming Interests are throttled in the similar way as that when there is only one victim server under attack.
Two servers' incoming traffic lines quickly go below the threshold after the attack started and soon merge the legitimate client traffic line, indicating that all the traffic received by the two servers are from legitimate clients.

\subsection{\sysname: Throttling Granularity}
\label{sec:evaluation:granularity}

In DDoS mitigation, collateral damage may ruin the legitimate traffic sent from the compromised devices.
\sysname throttles Interest traffic at a granularity of flow under a specific name prefix.
We evaluate \sysname in terms of the granularity of the traffic throttling.
We reuse the simulation settings of the two-victim scenario to let all the attackers not only attack $P$ with Valid-D Interests at 100 pkt/s but also keep the normal communication with another service provider $P'$ at the reasonable rate of 20 Interests/s.
In this case, the attacking traffic will overwhelm $P$ but the legitimate traffic will not go beyond the capacity of $P'$.
As shown in the Figure~\ref{fig:granularity-ddos}, compared with the two-victim scenario where traffic to both $P$ and $P'$ will be throttled, in this simulation, \sysname only squelches clients' traffic under $P$ while the traffic towards $P'$ will not be affected.

\subsection{\sysname: ``Smart" Attackers}
\label{sec:evaluation:smart}

Attackers may try to circumvent \sysname's traffic throttling by complying feedback messages from edge routers temporarily and switching back to attack mode later.
However, when attacks switch back to high sending rate, the traffic volume will alarm the service's pre-configured threshold again, which will in turn force the attackers to lower down the Interest sending rate again.
In this way, the \sysname will be triggered periodically when switches take place (Figure~\ref{fig:smart-attacker}).
Consequently, the average attacking traffic will be kept at a certain level where the service will not be spoiled, especially when the threshold is properly set below the real capacity of the service.

\subsection{\sysname: Mixed Interest Attack}
%We set $P$'s valid Interest capacity to be 500 in this simulation.

We tune the attackers in Valid Interest flooding scenario to send both Valid-D and Invalid Interest packets to simulate a mixed Interest attack.

\begin{figure} [ht]
	\centering
	\begin{subfigure}{.23\textwidth}
		\centering
		\includegraphics[height=4cm]{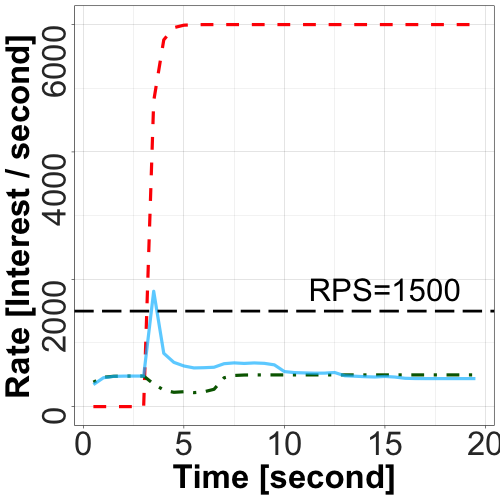}
		\caption{Mixed Interest DDoS}
		\label{fig:mixed-attack}
	\end{subfigure} %
	~
	\begin{subfigure}{.23\textwidth}
		\centering
		\includegraphics[height=4cm]{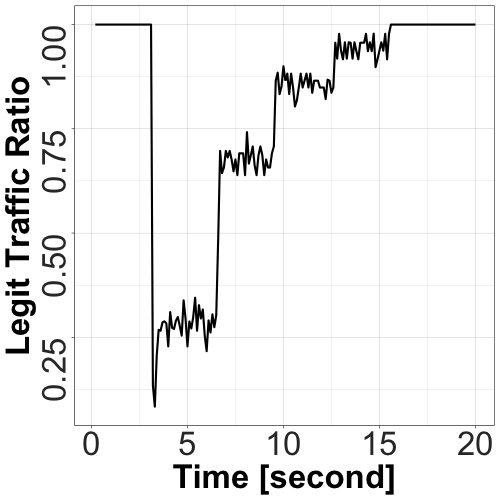}
		\caption{Legitimate Traffic Ratio}
		\label{fig:mixed-percentage}
	\end{subfigure}
	\vspace{2pt}
	{\footnotesize \\
	\textbf{Simulation Settings:}
	NDN deployment with \sysname.
	Each attacker (60 in total) starts sending both Invalid and Valid-D Interests at 100 pkt/s from second 3. Legitimate clients' sending rate is 40 pkt/s starting from second 0.
	We let $P$'s capacity be 1.5K RPS and \emph{RateLimitTimer} be 3 seconds.
	\par}
	\caption{\sysname: Mixed Interest Attack}
	\vspace{-0.5cm}
	\label{fig:mixed-ddos}
\end{figure}

As shown in Figure~\ref{fig:mixed-ddos}, compared with results of Valid-D Interest flooding scenario, one obvious difference is that, after the attack starts, \sysname will drop the traffic to be much lower than $P$'s \emph{RPS} (black horizontal line).
This is because at the edge, invalid Interest attack will lead to a total block of the attackers, which cancels out the $RPS$ from the valid Interest attack mitigation.
After a short period, \sysname will place the limit to misbehaving clients only and the legitimate clients will recover.
In the end, \sysname will only pass legitimate traffic to $P$ (Figure~\ref{fig:mixed-percentage}).
\section{Related Work}
\label{sec:related}
%There have been various solutions constructed around defending DDoS attacks.
%Following the taxonomy defined in the previous works~\cite{DDoStaxonomy, geva2014bandwidth}, we take several representative approaches as examples and analyze how those mechanisms work but are limited by the nature of IP networking.

There is a plethora of proposed DDoS defenses that operate at different levels and networking layers~\cite{bakr2019survey,vishwakarma2020survey}. Currently, deployable filter-based network-level remediation approaches in TCP/IP like FlowSpec\cite{RFC5575}, RTBH\cite{RFC5635}, the IETF's Distributed Open Threat Signaling (dots) Working Group\cite{ietf-dots}  lack the necessary expressiveness to address DDoS threats.  Effective DDoS remediation mechanisms generally must also depend on Deep Packet Inspection (DPI) to gain additional traffic insight which IP's stateless forwarding cannot provide. For instance, black-hole filtering blacklists entire network prefixes, which can cause collateral damage to: well-behaved sources, non-attack traffic that is sourced from compromised devices. As another example, FlowSpec requires the proper n-tuple consisting of several matching criteria so that DDoS traffic can be classified; however, since most supported matching criteria is at the network layer FlowSpec can mistakenly drop legitimate traffic to other services deployed on the same victim server or block good traffic sent from the compromised bots. Ioannidis and Bellovin proposed router-based Pushback~\cite{ioannidis2002pushback} which utilizes a heuristics function to detect packets that \emph{probably} belong to an attacker by checking the ``congestion signatures'' in the traffic. Due to routers' coarse-grained inspection of traffic, the filtering can lead to collateral damage.

Another interesting work is Active Internet Traffic Filtering (AITF)~\cite{argyraki2005active}, which requires the routers on the path of attacking traffic to mark traffic flows with route records (RRs), which are the IP addresses of routers who have forwarded the flow.
In this way, the victim can report the unexpected flows to the network, and routers can filter the flows identified by the RRs. However, this may potentially cause damage to
\begin{enumerate*} [label=(\roman*)]
	\item both attack and legitimate traffic because RRs cannot distinguish application-level flows,
	and
	\item benign hosts who are under the same first hop network as the attackers, because routers do not have fine-grain state to identify exact senders.
\end{enumerate*}. In \cite{Ang05} the authors proposed a DDoS defense using an overlay network without any architectural changes or client-side edge router enforcement.
%In addition, deployment of these approaches would require updating routing infrastructure in core Internet Service Provider (ISP) routers, without incentives for those ISPs to undertake such an upgrade.
Furthermore, StopIt~\cite{liu2008filter} and SIFF~\cite{yaar2004siff} require additional features that are missing in the existing TCP/IP architecture. Indeed, SIFF introduces privileged communications, which requires additional information carried by the IP header and each router on the connection path marking the IP packets.
%Filter-based and rate limiting approaches such as Ingress Filtering~\cite{RFC2827}, Pushback~\cite{ioannidis2002pushback}, Remote Triggered Back-Holing (RTBH)~\cite{RFC5635} and some other recent works~\cite{liu2008filter, yaar2004siff} utilize detection mechanisms to identify offending traffic and control the traffic by filtering it out or applying rate limiting.
%These mechanisms require Deep Packet Inspection (DPI) to have an insight into the ongoing traffic.
%However, the forwarding in IP routers lacks the state of the traffic; for example, the IP forwarder cannot distinguish the flows from different applications without additional overhead.
%Thus, these mechanisms take the risk of causing collateral damage on the legitimate traffic.
Capability-based approaches like TVA (Traffic Validation Architecture)~\cite{xiaowei05-dos-limiting, liu2008passport} introduce authentication of the packet source into the network system.
Taking TVA as an example, by embedding cryptographic authentication info into the IP packet, the routing system and servers are able to distinguish legitimate users from ``bad'' ones.
Operationally deploying these solutions requires adding extra functionality into the deployed TCP/IP architecture while causing incentive misalignment similar to Pushback.

%Ioannidis and Bellovin proposed a realization of router-based Pushback~\cite{ioannidis2002pushback} based on IP networking.
%To overcome the difficulty of knowing with certainty whether a packet actually belongs to a "good" or a "bad" flow, their proposed solution utilizes a heuristics function to detect packets that \emph{probably} belong to an attacker.
%This is done by using the ``congestion signature".
%Routers can thus preferentially drop the packets.
%However, it is hard to implement this Pushback defense, due to the difficulty of getting the congestion signature and the lack of forwarding insights provided by the routers.

%Compared with \cite{ioannidis2002pushback}, our approach based on NDN utilizes the victim's feedback and the traffic states provided by NDN at each router would help the system know with certainty the source of the overwhelming traffic.
%Ultimately, our proposed Pushback learns which clients are not obeying the DDoS control and has the ability to selectively rate limit only those clients.
%% <EO> I don't know if we want to broach this, but it is part of our story...
% Additionally, these approaches (Pushback, TVA, etc.) lack an incentive model for incremental deployment.
%% </EO>

Due to strict layering, existing DDoS mitigation solutions in TCP/IP networks have faced difficulties when trying to get better insights of the traffic flows or to modify the architecture. In contrast, NDN enables \sysname to identify and throttle specific application-level DDoS traffic flows at fine granularity by leveraging NDN's architectural features. As such, victim is able to report the DDoS at name prefix granularity, and \sysname is able to identify exact application-level attack traffic flows and attackers at the network level.

%\begin{enumerate*} [label=(\roman*)]
%	\item base DDoS traffic signature on matching criteria from network headers (e.g., IETF dots, Pushback, BGP FlowSpec, RTBH)
%	or
%	\item add additional features to the network architecture for finer flow classification (e.g., AITF, SIFF, StopIt, TVA).
%\end{enumerate*}
%Due to strict layering, identifying specific application-level traffic flows is unattainable. Consequently, these solutions have faced difficulties in real world deployment and collateral damage caused by the coarse granularity. By contrast, NDN enables \sysname to identify and throttle specific application-level DDoS traffic flows at fine granularity by leveraging NDN's architectural features:
%\begin{enumerate*} [label=(\roman*)]
%	\item application-semantic meaningful names
%	and
%	\item stateful forwarding
%\end{enumerate*},
%(as shown in the simulation scenario in~\ref{sec:evaluation:granularity}).

\subsection{Related Works in NDN/ICN}
\label{sec:ndn:previous}

There have also been various proposed approaches to mitigate Interest DDoS over NDN/ICN~\cite{zhang2014named,CCN}.
Specifically, the authors of work \cite{afanasyev2013interest, compagno2013poseidon} leverage the ``success ratio" (how many Interests get satisfied by Data) to detect the presence of fake Interest DDoS.
Others~\cite{dai2013mitigate, salah2016evaluating} propose to detect Interest flooding by monitoring the PIT size or PIT utilization rate. They mainly focus on one specific type of attack -- Invalid Interest attack (i.e. Interests carrying non-existing data names).
To be effective, they also require that routers must be able to set proper threshold values for the detection function, and these threshold values can be non-trivial to configure when underlying traffic composition is complex. In comparison, \sysname directly takes input from attack targets and can handle valid Interest (i.e. Valid-S and Valid-D) flooding and mixed Interest attack scenarios where attackers can all types of attack Interests towards the target. The explicit feedback from the victim enables accurate traffic throttling: reinforcement throttling focuses the enforcement only to misbehaving clients, removing the need to configure proper threshold values.
\section{Conclusion}
\label{sec:conclusion}

% The only remaining volume-style of network attack is Interest flooding, which is mitigated by our defense model, PAP.
% By reviewing existing DDoS attack types for IP and potential flooding attacks over NDN, we learned that even the already DDoS-resistant NDN needs safeguards against possible abuses.
% Therefore, by utilizing NDN's stateful forwarding and structured names, we developed a system that is capable of actively responding to all three types of Interest flooding.
%%
%IP is being abused to facilitate distributed attacks,
%our mitigation techniques and defenses have struggled with fundamental misalignments between the essential functions and forwarding semantics needed for effective mitigation and IP's stateless forwarding, to address and remediate the traffic from large DDoS attacks. Although many solutions have been proposed, misalignments of incentives and cost/benefit tradeoffs make their rollout difficult. 
%Consequently, the current mitigation techniques necessarily backhaul DDoS attack traffic across the Internet to centralized mitigation servers that do DPI.
%This results in congesting links, costing operational overhead, and framing an unmaintainable capacity mismatch (in which transit capacity of centralized mitigation must match the aggregate DDoS traffic from distributed attack sources).  %The ability to remediate DDoS attacks close to their sources has been among the goals that have long been sought after.

%% 1/ The vulnerabilities at the core of the TCP/IP architecture enable DDoS %% 2/ stateless IP makes countermeasures difficult to implement.

DDoS attacks have been an asymmetric threat since they first became significant more than 21 years ago.
The asymmetry exhibits in two ways: an attack costs nothing to launch but can cause multi-million dollar losses to victims;
and an attack is trivial to launch but extremely difficult to defend.
Previous work~\cite{jin2003hop,off-by-default,xiaowei05-dos-limiting,handley04-steps,rossow2014amplification} attributed the first one to IP's original design goal of maximizing reachability by allowing any host to send packets to any other host without considering security.
In this paper we attribute the second asymmetry to IP's semantic-free addresses and stateless forwarding plane. This makes traffic flooding easy  - but adding floodgates is difficult - as floodgates represent an architectural change and a misalignment of incentives in general.

%% 3. Instead of continually patch extra functionalities to the current TCP/IP architecture, we explore a DDoS-resilient architecture
%% 4/ network can do more if given the needed functions

Recognizing the above limitations, in this paper we take a radically new direction to search for effective DDoS mitigation solutions.
In addition to having some of the most desired DDoS resiliency properties natively built-in (separation of clients and servers, symmetric traffic, etc.), the NDN design provides two key enabling features for DDoS mitigation: use of semantic names at the network layer and stateful forwarding.
The former enables fine-grained traffic identification (that DPI attempted to achieve), and the latter enables throttling attack traffic along the paths all the way to the sources (that many of the previous efforts had aimed to achieve~\cite{bakr2019survey,vishwakarma2020survey,ietf-dots,RFC5575,RFC5635,ioannidis2002pushback}).
We propose \sysname to demonstrate how new architectural features can lead to effective solutions to DDoS mitigation.  We show that the network can indeed be expected to do more, once we equip it with the necessary functionality.
 
Given every coin has two sides, the road to FITT/NDN deployment also has its cost. As a new architecture, NDN needs to be deployed at end nodes as well.  Although we identified a path to incremental deployment using CDNs, FITT also requires the adoption of NDN gateways at the client and operator sides.
%% In addition, NDN  can expose app data names which can be viewed as a potential privacy threat.
% Nevertheless given the stagnant progress in DDoS mitigation in decades, 
Nevertheless, with DDoS mitigation losing ground to increasingly large and complex attacks, we hope a radical new solution can help shed new insight on the solution space.

\bibliographystyle{IEEEtran}
\bibliography{defense}

\end{document}